\documentclass[useAMS,usenatbib]{mn2e}

\newcommand{\araa}{{ARA\&A}}          
\newcommand{\apj}{{ApJ}}         
\newcommand{\apjl}{{ApJ}}          
\newcommand{\aap}{{A\&A}}          
\newcommand{\mnras}{{MNRAS}}          
\newcommand{\nat}{Nature}       

\newcommand{\iaucirc}{IAU~Circ.}
\newcommand{\physrep}{Physics Reports}

\usepackage{amssymb}
\usepackage{amsmath}
\usepackage[dvips]{graphicx}
\usepackage{epsfig}
\usepackage{natbib}
\usepackage{multicol}
\usepackage{rotating}
\usepackage{booktabs}
\usepackage{caption}
\usepackage{tikz}
\usetikzlibrary{shapes,arrows}

\label{firstpage}
\date{Accepted ... Received ...; in original form ...}
\pagerange{\pageref{firstpage}--\pageref{lastpage}} \pubyear{2012}
\title[Flux prescriptions for gamma-ray burst afterglows]{Practical flux prescriptions for gamma-ray burst afterglows, from early to late times}
\author[Leventis et al.]{K. Leventis$^{1}$\thanks{E-mail: K.Leventis@uva.nl}, H. J. van Eerten$^{2}$, Z. Meliani$^3$, R.A.M.J. Wijers$^1$\\
$^{1}$Astronomical Institute `Anton Pannekoek', PO box 94248, 1090 SJ Amsterdam, the Netherlands\\
$^{2}$Center for Cosmology and Particle Physics, Physics Department, New York University, New York, NY 10003, USA\\
$^{3}$Laboratoire Univers et Th{\'e}ories, Observatoire de Paris, UMR 8102 du CNRS, Universit{\'e} Paris Diderot, 92190, Meudon, France}

\begin{document}

\maketitle

\begin{abstract}
We present analytic flux prescriptions for broadband spectra of self-absorbed and optically thin synchrotron radiation from gamma-ray burst afterglows, based on one-dimensional relativistic hydrodynamic simulations. By treating the evolution of critical spectrum parameters as a power-law break between the ultrarelativistic and non-relativistic asymptotic solutions, we generalize the prescriptions to any observer time. Our aim is to provide a set of formulas that constitutes a useful tool for accurate fitting of model-parameters to observational data, regardless of the dynamical phase of the outflow. The applicability range is not confined to gamma-ray burst afterglows, but includes all spherical outflows (also jets before the jet-break) that produce synchrotron radiation as they adiabatically decelerate in a cold, power-law medium. We test the accuracy of the prescriptions and show that numerical evidence suggests that typical relative errors in the derivation of physical quantities are about 10 per cent. A software implementation of the presented flux prescriptions combined with a fitting code is freely available on request and on-line \footnotemark. Together they can be used in order to directly fit model parameters to data.

\end{abstract}


\vspace{1cm}


\section{Introduction}\label{intro}
\let\thefootnote\relax\footnotetext{$\!\!\!\! \dag \,\,$ The URL is http://www.astro.uva.nl/research/cosmics/gamma-ray-bursts/software/.}
Gamma-ray bursts (GRBs) are believed to be produced by powerful relativistic outflows resulting from the catastrophic death of massive stars (\citealt{Woosley1993}), or the merger of two compact objects (\citealt{Eichler1989}). The burst itself (prompt emission) likely arises from internal shocks occurring due to the variability of the central engine \citep{Rees1994,SariPiran1997}, while the \textit{afterglow} emission comes from the interaction of the same outflow with the medium surrounding the burster \citep{Rees1992,Paczynski1993}. Although the dominant radiation process behind the prompt emission is not yet clear, it is well established that the afterglow radiation is dominated by synchrotron emission from shock-accelerated electrons \citep{MeszarosRees1993,vanParadijs2000}.

The prompt emission is typically very brief and concentrated at high energies. On the other hand, afterglows are often visible over many more orders of magnitude both in time- and frequency-space (see \citealt{Meszaros2006} for an extensive review of GRB research). Thus, studying the afterglow radiation allows us to put a multitude of constrains both on the microphysics (e.g. the fraction of internal energy going to the magnetic fields and the power-law accelerated electrons) governing the shocked plasma \citep{Spitkovsky2008,Sironi2009}, as well as on the basic physical parameters describing the phenomenon macroscopically, like blast-wave energy, density and structure of the surrounding medium.

It is these macroscopic parameters that determine the dynamical evolution of the outflow. However, a full analytic description of the dynamics is only possible when the spatial component of the four-velocity of the outflow $\beta \gamma$ is either much greater \citep{Blandford1976} or much smaller \citep{Sedov1959} than $1$. Therefore, relativistic hydrodynamic (RHD) simulations \citep{Kobayashi1999,Meliani2007,Zhang2009,DeColle2012a} are the most accurate means of studying the intermediate dynamical regime linking the ultrarelativistic and Newtonian solutions (see however \citealt{Huang1999}). Van Eerten et al. (2010a) have numerically studied the lightcurves of outflows advancing through all three dynamical regimes and have shown that the transition is slow, i.e. deviations from the expected relativistic behaviour appear well before the Newtonian asymptotes are reached, mainly due to the changing adiabatic index of the shocked gas. For typical burst parameters (isotropic blast-wave energy $E_{\textrm{iso}}=10^{52}\, \textrm{ergs}$, ambient medium number density $n_{0}=1\, \textrm{cm}^{-3}$) the Sedov-Taylor scalings set in at a few thousand days, observer time, implying that an appreciable portion of the afterglow (typically around hundreds of days) emanates from outflows with dynamics that cannot be described analytically by either of the two asymptotic solutions. 

Soon after the discovery of the first afterglows \citep{Costa1997,Groot1997} efforts were made to calculate broadband synchrotron spectra and light curves as a function of burst parameters \citep{Wijers1997,Sari1998,Panaitescu2000}. The common way to do this is by tying the dynamical evolution of the blast-wave in regimes where this is feasible to radiation models that, according to the jump conditions at the shock front, calculate the resulting spectra. Despite the success of early efforts in capturing general features of the observed spectra, the progressive refinement of the used models has led to very different estimates of the physical parameters of individual bursts. For example \citet{Wijers1999} and \citet{Granot2002} have both fitted GRB 970508 and their derived values differ up to 3 orders of magnitude. Furthermore, the applicability of most of these models is restricted to a particular dynamical phase and only recently have there been a few attempts at addressing the entire evolution of spectra and light curves through the performance of simulations \citep{Zhang2009, vanEerten2012b,Wygoda2011, DeColle2012a}. Even so, these models do not contain a treatment of self-absorption (apart from \citealt{vanEerten2012b}), necessary to model low-frequency observations with e.g. the Expanded Very Large Array EVLA (\citealt{Perley2011}), the Low-Frequency Array LOFAR (\citealt{Morganti2011}) and the upcoming Karoo Array Telescope MeerKAT (\citealt{Booth2009}) and Square Kilometre Array SKA (\citealt{Carilli2004}), and do not provide flux-prescriptions. Van Eerten et al. (2012) do provide a broadband fit code, but it requires the use of a parallel computer network.

The purpose of this work is to provide accurate analytic flux-prescriptions, based on one-dimensional RHD simulations, that are applicable to both the ultrarelativistic and Newtonian phase but also, and perhaps more importantly, to observer times when the outflow is transitioning from the former to the latter. Apart from the typical, initially ultrarelativistic outflows of GRBs, the formulas we present are applicable to Newtonian as well as relativistic (\citealt{Soderberg2010}) outflows from supernova explosions in the adiabatic phase \citep{Chevalier1977,Chevalier1982,Draine1993} and mildly relativistic outflows originating from binary neutron star (NS) mergers, expected to produce detectable electromagnetic (EM) counterparts to gravitational wave detections \citep{Nakar2011, Metzger2012b}. They can also be applied to relativistic outflows resulting from the tidal disruption of stars by a super-massive black hole \citep{Bloom2011,Metzger2012a}, under the limiting assumption of quasi-spherical outflow. The presented model naturally accounts for the exact shape of the synchrotron spectrum (including self-absorption, but ignoring cooling) and the structure of the blast-wave. Furthermore, it can be applied to a range of power-law density structures of the circumburst medium, a possibility previously studied by \citet{vanEerten2009} and \citet{DeColle2012b}. This allows for modelling of more complex environments, expected on a theoretical basis (\citealt{Ramirez-Ruiz2005}) and deduced observationally (\citealt{Curran2009}). With such a tool a light curve can be fitted without the need of costly simulations and the restrictions of models specialising in specific dynamical phases, or preset structures of the circumburst medium. In order to obtain the flux-prescriptions we combine three elements: (1) analytic formulas for flux-scalings during the Blandford-McKee and the Sedov-Taylor phases, (2) one-dimensional, hydrodynamic simulations, using the adaptive mesh refinement code \textsc{amrvac} \citep{Meliani2007,Keppens2012}, that span the whole range of the dynamics (from ultrarelativistic to Newtonian velocities) and (3) a radiative-transfer code that uses simulation snapshots and a parametrisation of the microphysics to calculate instantaneous spectra.

This paper is organised as follows: in section \ref{numerics} we briefly describe the setup of the performed simulations and the subsequent calculations of spectra and light curves. In section \ref{flux-prescriptions} we present formulas that describe the flux as a function of physical parameters in both the relativistic and the Newtonian phase of the outflow. That includes specifying the flux at any given power-law segment, as well as a description of the sharpness of the spectral breaks that occur at critical frequencies. We then proceed in section \ref{transrel} to connect the two dynamical regimes (relativistic and Newtonian) by treating the transition from the former to the latter as a prolonged temporal \textit{break} the characteristics of which can be linked to the physical parameters of the burst and its environment. In section \ref{using} we describe how one can make use of the flux-prescriptions to obtain spectra at any given time. We also show comparisons between spectra based on simulations and spectra constructed using the provided prescriptions. Finally, we present an application of this model to mildly relativistic outflows from binary neutron star mergers in order to assess the recent predictions of \citet{Nakar2011} concerning the detectability of the produced radio signals. In section \ref{discuss} we discuss our results and the implications of this work for GRB afterglow models.


\section{Numerical treatment}\label{numerics}

\subsection{Simulations}
We have made use of the \textsc{amrvac} adaptive-mesh-refinement numerical code to run a series of simulations for different values of physical parameters. These simulations span a wide range of the four-velocity at the shock front, from ultrarelativistic values ($\sim70$) down to $\sim0.05$.

In total 7 different simulation runs were used to arrive at the presented prescriptions. They can be characterized by the blast-wave energy $E_{52}$ (in units of $10^{52}\, \textrm{erg}$), starting Lorentz factor of the shock $\Gamma_{\textrm{in}}$, maximum radius of the simulation-box $R_{\textrm{max}}$, slope of the power-law density distribution of the surrounding medium $k$ and value of the number density at $10^{17}\, \textrm{cm}$, $n_0$. In Table \ref{Sim-tab} we present the values of these parameters for each run.

\begin{table}
\centering
\parbox{0.85\columnwidth}{\captionof{table}{Parameters of simulations used to derive the flux prescriptions. Left column enumerates the performed simulations.\label{Sim-tab}}}
\small\addtolength{\tabcolsep}{+2pt}
\begin{tabular}{l c c c c c} 
\toprule                        
 Sim & $E_{52}$ & $k$ & $n_0$ & $\Gamma_{\textrm{in}}$ & $R_{\textrm{max}}\, (\textrm{cm})$ \\ [0.5ex] 
\hline                  
$1$ & $1.0$ & $0.0$ & $1.0$ & $60$ & $3\cdot10^{19}$  \\ [1.5ex]
$2$ & $0.04$ & $0.0$ & $4.0$ & $60$ & $10^{19}$  \\ [1.5ex]
$3$ & $1.0$ & $0.5$ & $1.0$ & $60$ & $4\cdot10^{19}$  \\ [1.5ex]
$4$ & $0.4$ & $0.75$ & $0.5$ & $28$ & $5\cdot10^{19}$  \\ [1.5ex]
$5$ & $1.0$ & $1.0$ & $1.0$ & $60$ & $4\cdot10^{19}$  \\ [1.5ex]
$6$ & $1.0$ & $2.0$ & $1.0$ & $70$ & $10^{19}$  \\ [1.5ex]
$7$ & $0.01$ & $2.0$ & $1.0$ & $10$ & $7\cdot10^{19}$  \\ [1ex]      
\bottomrule       
\end{tabular}\par
\smallskip
\end{table}

\subsubsection{Resolution}
The 1D simulation box typically extends from $10^{16}\, \textrm{cm}$ up to $\textrm{a few times}\, 10^{19}\, \textrm{cm}$, although both limits were modified accordingly for physical models with different external density profiles and blast-wave energies. Utilizing the adaptive-mesh-refinement approach in \textsc{amrvac} we have used a maximum of 20 refinement levels which set the effective resolution of the grid. For 120 cells at the lowest refinement level, this amounts to a resolution of $\sim 4.77\cdot 10^{11}\, \textrm{cm}$ per cell. For comparison, the corresponding width of the initial BM shell for Simulation 1 is (\citealt{vanEerten2010b}) $R_{\textrm{in}}/(6\Gamma_{\textrm{in}}^2) = 5\cdot 10^{12}\, \textrm{cm}$, with $R_{\textrm{in}}$ denoting the radius of the shock.

We have checked for convergence against runs of different refinement levels (see also \citealt{vanEerten2011a}, \citealt{DeColle2012a}) and have also checked our results against theoretically predicted values in the early part of the outflow when the resolution demands are the highest. They have been found in good agreement.

\subsubsection{Equation of state}
In all the simulations we have used a `realistic' equation of state (EOS) with an effective adiabatic index (\citealt{Meliani2004}) that lies between the ultrarelativistic and non-relativistic limits ($4/3$ and $5/3$, respectively):

\begin{equation}
\Gamma_{\textrm{ad,eff}} = \frac{5}{3} - \frac{1}{3} \left( 1 - \frac{\rho'^2 c^4}{u'^2} \right).
\end{equation} 

In the above equation $\rho' c^2$ is the comoving rest-mass energy density and is weighed against the total energy density (including rest-mass) of the gas $u'$.

This Synge-type (\citealt{Synge1957}) EOS has also been used in \citet{vanEerten2010a}, where the effects have been analysed and comparisons to constant $\Gamma_{\textrm{ad,eff}}$ have been made. In short, its effect on the observed flux is that of a very gradual transition from values close to (but not at) those corresponding to an ultrarelativistic EOS to values approaching those of a non-relativistic EOS.

\subsection{Radiative-transfer code}

The snapshots generated by the simulation runs were post-processed using a radiative-transfer code \citep{vanEerten2009,vanEerten2010a}. During the post-processing an array of beams is created and propagated at the speed of light through the three-dimensional generalisation of the 1D snapshots towards the observer. The elements of this array take the value of the specific intensity $I_{\nu}$. At each step the solution to the equation of radiative transfer is applied to each beam and the value for the intensity is updated through the equation

\begin{equation}
I_{\nu} = I_{0} \, \textrm{e}^{-\tau_{\nu}} + S_{\nu} \, ( 1 - \textrm{e}^{-\tau_{\nu}} ) ,
\end{equation}
where $I_{0}$ is the value of the intensity at the previous step, $\tau_{\nu}$ is the optical depth and $S_{\nu}$ the source function. We note that the optical depth of a single simulation-cell can be larger than unity.

The array of intensities is constructed so that the positions of its elements lie on the surface from which light-signals arrive at the observer simultaneously. The density of the beams is determined through an adaptive-mesh approach ensuring sufficient resolution. Once all beams have crossed the entire blast-wave integration of the intensity over the surface yields the flux

\begin{equation}\label{integral-intens}
F_{\nu} = \frac{(1+z)}{d^2} \int \limits_A \, I_{\nu} \, d\!A ,
\end{equation}
where $d$ is the luminosity distance to the observer, $(1+z)$ the cosmological correction and $A$ the surface defined by the beams.

In the case of on-axis jets and spherical outflows, like the ones considered in this study, eq. (\ref{integral-intens}) can be reduced to a 1D integral due to the axisymmetry of the source (\citealt{GPS1999}). Our approach makes use of that symmetry and a 1D integral is solved numerically to calculate the observed flux.


\section{Flux-prescriptions in the asymptotic dynamical regimes}\label{flux-prescriptions}

In this section we demonstrate how to combine blast-wave dynamics in each of the self-similar regimes with synchrotron radiation theory to arive at scalings that describe the observed flux as a function of frequency and time.

\subsection{Shock dynamics}

We first outline the dynamics of the outflow. As mentioned in section \ref{intro}, one can obtain power-law scalings for the four-velocity as well as the mass and energy densities right behind the shock front as a function of blast-wave radius and time in the two extreme regimes of dynamical behaviour of the afterglow.

\subsubsection{Ultrarelativistic phase}
In the ultrarelativistic (also known as Blandford-McKee, hereafter BM) phase these scalings take the form (\citealt{Blandford1976}):

\begin{equation}\label{gamma-BM}
\gamma_{\textrm{2}}\propto \Gamma_{\textrm{sh}} \propto t^{-\frac{3-k}{2}},
\end{equation}
\begin{equation}
e_{\textrm{2}}\propto \Gamma_{\textrm{sh}}^2\, n_1,
\end{equation}
\begin{equation}
n_{\textrm{2}}\propto \Gamma_{\textrm{sh}}\, n_1.
\end{equation}
where $\gamma_{\textrm{2}}$ is the Lorentz factor of the shocked plasma measured in the lab frame (which coincides with the frame of the surrounding medium), $e_{\textrm{2}}$ describes its internal energy density and $n_{\textrm{2}}$ its number density. Here and throughout this paper the quantities $e$ and $n$ will be measured in the comoving frame of the fluid they are describing. $\Gamma_{\textrm{sh}}$ is the Lorentz factor of the propagating shock wave, while $n_1 (r) = n_0\, (r/r_0)^{-k}$ is the density of the unshocked medium surrounding the burster as a function of radius. In all the results presented the characteristic distance $r_0$ is put at $10^{17}\, \textrm{cm}$. The time `$t$' appearing in eq. (\ref{gamma-BM}) is the lab-frame time and is to be distinguished from the observed arrival time of light signals which is affected by light travel-time effects. While in the BM phase we can assume $r\sim c\, t$ for the radius of the shock.

\subsubsection{Newtonian phase}

At the phase where the outflow has become Newtonian (also known as Sedov-Taylor phase, hereafter ST) a similar approach can be taken to describe its kinetic and thermodynamic evolution. Dimensional analysis implies

\begin{equation}
r(t) \propto \left( \frac{E\, t^2}{n_1(r)} \right) ^{1/5}
\end{equation}
for the scaling of the radius of the shock as a function of time, blast-wave energy and structure of the surrounding medium. This leads to

\begin{equation}
\beta(t) \propto t^{-\frac{3-k}{5-k}},
\end{equation}
\begin{equation}
e_2 \propto t^{\frac{-6}{5-k}},
\end{equation}
\begin{equation}\label{n_2ST}
n_2 \propto n_1,
\end{equation}
where $\beta = \dfrac{dr}{c\,dt}$ is the bulk velocity of the shock, in units of $c$.

\subsection{Optically thin and self-absorbed synchrotron radiation}

The next step is to combine these scalings with formulas that calculate optically thin as well as self-absorbed synchrotron radiation. We assume that electrons are the primary radiating particles and express their post-shock energy distribution as a power-law $N(E) \propto E^{-p}$. The lower limit of the distribution $E_{\textrm{m}}=\gamma_{\textrm{m}}\, m_{\textrm{e}} c^2$ corresponds to a comoving synchrotron frequency $\nu_{\textrm{m}}'=\dfrac{3}{4\pi}\gamma_{\textrm{m}}^2\dfrac{Q_{\textrm{e}}\, B'}{m_{\textrm{e}}\, c}\textrm{sin}\alpha$, where $Q_{\textrm{e}}$ and $m_{\textrm{e}}$ are the electron charge and mass, respectively, $B'$ is the comoving value of the magnetic field and $\alpha$ is the pitch angle between magnetic field and velocity of the electron. In the case of optically thin radiation the flux at a given frequency $\nu'$, in the comoving frame will be

\begin{equation}\label{F-thin}
F'_{\nu'} \propto (p-1)\, N \, B' \, \mathcal{Q}(y_\textrm{m}),
\end{equation}
where $N$ is the total number of power-law accelerated electrons, $y_{\textrm{m}}=\dfrac{\nu'}{\nu_{\textrm{m}}'}$ and
\begin{equation}\label{Q(x)}
\mathcal{Q}(x) \equiv x^{\frac{1-p}{2}}\, \int_0^x y^{\frac{p-3}{2}}\,\mathcal{P} (y) \, \textrm{d}y.
\end{equation}
The function $\mathcal{Q}$ contains all the spectral information. The function $\mathcal{P}$ appearing in eq. (\ref{Q(x)}) is the synchrotron function $F(x)$ (\citealt{Rybicki1986}) integrated over all pitch angles, for an isotropic pitch-angle distribution. 

\begin{figure}
\includegraphics[width=\columnwidth]{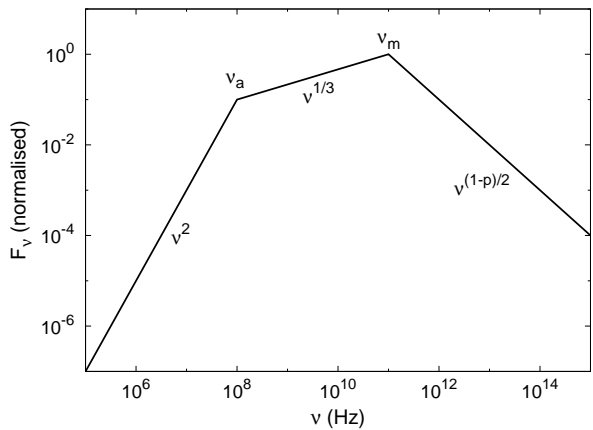}
\caption{Spectrum 1. Normalised form of the spectrum when $\nu_{\textrm{a}} < \nu_{\textrm{m}}$.}
\label{fig_spec_am}
\end{figure}

\begin{figure}
\includegraphics[width=\columnwidth]{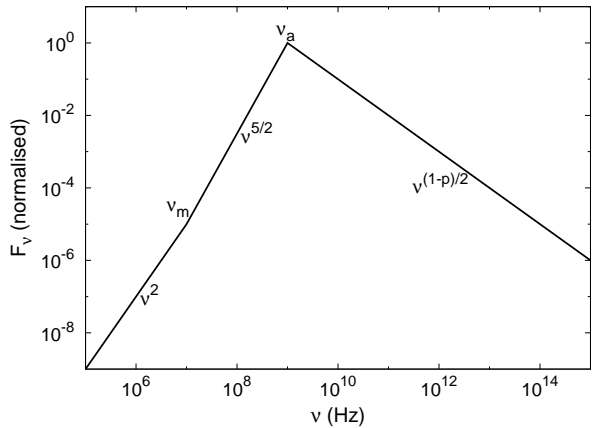}
\caption{Spectrum 2. Normalised form of the spectrum when $\nu_{\textrm{m}} < \nu_{\textrm{a}}$.}
\label{fig_spec_ma}
\end{figure}

In the case of optically thick synchrotron radiation the comoving flux is given by

\begin{equation}
F'_{\nu} \propto \frac{j'_{\nu}}{\alpha'_{\nu}} \, r'^2,
\end{equation}
where $j'_{\nu}$ and $\alpha'_{\nu}$ are the comoving emissivity and absorption coefficient, respectively, and $r^2$ is a measure of the radiating surface. The expressions for $j'_{\nu}$ and $\alpha'_{\nu}$ have the form

\begin{equation}
j'_{\nu} \propto (p-1)\, \xi \, n_2 \, B' \, \mathcal{Q}(y_{\textrm{m}}),
\end{equation}
\begin{equation}\label{abs-eq}
\alpha'_{\nu} \! \propto \! \frac{(p-1)^2(p+2)}{p-2} \, \xi^2 \, n_2^2 \, \epsilon_{\textrm{e}}^{-1} \, e_{2}^{-1} \, B' \, \nu'^{-2} \, \mathcal{Q}(p+\!1,y_{\textrm{m}}),
\end{equation}
where $\xi$ is the fractional number of electrons accelerated to a power-law distribution, $\epsilon_{\textrm{e}}$ is the fraction of internal energy carried by the accelerated electrons and $\mathcal{Q}(p+1,y_{\textrm{m}})$ is evaluated using eq. (\ref{Q(x)}) by replacing $p$ with $(p+1)$. The effect of absorption is the introduction of another critical frequency in the spectrum, $\nu_{\textrm{a}}$. The ordering of $\nu_{\textrm{m}}$ and $\nu_{\textrm{a}}$ determines the shape of the spectrum. In Fig. \ref{fig_spec_am} and \ref{fig_spec_ma} the two different spectra are represented schematically in order to illustrate the break frequencies as well as the slopes of the power laws that they connect.

To arrive at the relations above we have demanded that the distribution of the electrons obeys

\begin{equation}
\int_{E_{\textrm{m}}}^{\infty} N(E)\,dE = \xi \, n_2
\end{equation}
and
\begin{equation}
\int_{E_{\textrm{m}}}^{\infty} N(E)\, E \, dE = \epsilon_{\textrm{e}} \, e_2,
\end{equation}
where we have used the implicit condition that $p > 2$.

In this study we ignore the effect of electron-cooling on the spectra since the fast timescales associated with it translate to distances much shorter than the typical size of a simulation-cell. This work focuses on observer times when the influence of cooling on the observed spectra is negligible.

\subsection{General form of flux-scalings}

Equations (\ref{gamma-BM})-(\ref{n_2ST}) allow us to calculate the conditions right behind the shock front as a function of time. From these equations we can compute instantaneous spectra by utilizing standard formulas for synchrotron radiation. However, eq. (\ref{gamma-BM})-(\ref{n_2ST}) do not specify the structure of the shocked plasma well behind the shock. Such a specification would have allowed us to convolve the different parts of the outflow that contribute to the observed radiation at any given observer time. Such an approach has been taken, for example, by \citet{Granot2002}.

Our approach is based on the fact that in the self-similar regimes the scaling-behaviour of the emitted radiation can be calculated by considering a homogeneous slab that obeys the scalings of the shocked fluid right behind the shock. However, in order to correctly calibrate the scalings (i.e. provide the correct flux-levels) one has to capture the shock structure behind the front and the most reliable way to do this is by simulations.

The calibration is done by introducing a polynomial in terms of $p$ and $k$ -- the spectral index of the power-law accelerated electrons and the index describing the structure of the surrounding medium, respectively. The former quantity ($p$) determines the electron distribution, everywhere behind the shock front, for a given set of thermodynamic parameters, while the latter ($k$) affects the structure of the decelerating blast-wave. Two standard values for $k$ are often assumed in the literature, namely $0$ (constant density medium) and $2$ (constant stellar wind environment). However, fits to $k$ \citep{Yost2003,Curran2009} often indicate different conditions, motivating us to use it as a free parameter. Values for the factors of the calibrating polynomial are then derived by demanding that they satisfy the system of equations resulting from runs of models with different $p$ and $k$. The range of values we have explored are $[2.1,3]$ for $p$ and $[0,2]$ for $k$. Therefore this is also the range under which the presented prescriptions are applicable.

Having put all of the ingredients together, the equation describing the flux at any given power-law segment of the synchrotron spectrum, either in the BM or the ST phase, has the general form

\begin{equation}\label{F_nu}
F_{\nu} = C_{\!\textrm{pol}}\, h(p)\, \xi^{q_{\xi}} \, \epsilon_{\textrm{e}}^{q_e} \, \epsilon_{\textrm{B}}^{q_B} \, n_0^{q_n} \, E_{52}^{q_E} \, \nu_{\textrm{obs}}^{q_{\nu}} \, t_{\textrm{obs}}^{q_t} \, (1\!+\!z)^{q_z} d_{28}^{-2},
\end{equation}
where
\begin{equation}\label{C_pol}
\textrm{log} \, C_{\!\textrm{pol}}= g_0 + g_p \, p + g_{pp} \, p^2 + g_k \, k + g_{kk} \, k^2
\end{equation}
and $h(p)$ is a function of $p$, different for each power-law segment. It takes the following values:

\begin{equation}\label{h(p)}
h(p) =
\begin{cases}
\vspace{2mm}
 \dfrac{(p-2)}{(p-1)(p+2)}\,\dfrac{3p+2}{3p-1}, & F_{\nu} \propto \nu^{2} \\
\vspace{2mm}
 \dfrac{1}{(p+2)}\,\dfrac{G(p)}{G(p\!+\!1)}, & F_{\nu} \propto \nu^{5\!/\!2} \\
\vspace{2mm}
\dfrac{p-1}{3p-1}\, \left(\dfrac{p-2}{p-1}\right)^{-2\!/\!3}, & F_{\nu} \propto \nu^{1\!/\!3} \\
\vspace{2mm}
(p-1)\, G(p)\, \left(\dfrac{p-2}{p-1}\right)^{p\!-\!1}, & F_{\nu} \propto \nu^{(1\!-\!p)\!/\!2}
\end{cases}
\end{equation}
where
\begin{equation}
G(p)=\frac{\Gamma\!\!\left(\frac{5}{4}+\frac{p}{4}\right) \, \Gamma\!\! \left( \frac{p}{4}+\frac{19}{12} \right) \, \Gamma\!\! \left( \frac{p}{4}-\frac{1}{12} \right)}{ \Gamma\!\! \left(\frac{7}{4}+\frac{p}{4}\right) \, (p+1)}.
\end{equation}
$G(p)$ as well as other factors appearing in eq. (\ref{h(p)}) originate from the limiting behaviour of $\mathcal{Q}(x)$. For details see \citet{vanEerten2009}.

Equation (\ref{F_nu}) shows all the possible physical dependencies of the flux that this model is taking into account. We have introduced the fraction of internal energy carried by the magnetic field $\epsilon_{\textrm{B}}$. This quantity, along with $\epsilon_{\textrm{e}}$, $\xi$ and $p$ constitute a group describing the microphysics of the shocked electrons and enter via synchrotron theory. Therefore, the exponents $q_{\xi}$, $q_e$, $q_B$ and the prefactor $h(p)$ remain the same regardless of the dynamics of the outflow. On the other hand there are two quantities describing the burster and its environment: $n_0$ and the blast-wave energy $E_{52}$ (measured in units of $10^{52}\, \textrm{erg}$), two quantities describing the frequency ($\nu_\textrm{obs}$) and the time ($t_{\textrm{obs}}$) of the observation and two more describing the cosmological distances usually associated with GRBs: the redshift $z$ and the luminosity distance $d_{28}$ (in units of $10^{28}\, \textrm{cm}$).

We note that the inclusion of $\xi$ in our description of the microphysics has the implication that one cannot uniquely determine the values of all model parameters at once. This is a consequence of the degeneracy of the used model which for a set of primed parameters $E_{52}'=E_{52}/f,\, n_0'=n_0/f,\, \epsilon_{\textrm{e}}'=f \epsilon_{\textrm{e}},\, \epsilon_{\textrm{B}}'=f \epsilon_{\textrm{B}},\, \xi'=f \xi$, produces the same spectrum as the set of unprimed parameters. This degeneracy was first pointed out by \citet{Eichler2005} and can also be seen in eq. (\ref{FmBM}), (\ref{FmST}) and Tables \ref{nu_cr-BM-am}-\ref{nu_cr-ST-ma} presented in Section \ref{transrel}. As a result, a value for one of the parameters must be assumed during fitting in order to determine the others.

All the $q$-exponents appearing in eq. (\ref{F_nu}) are determined analytically. They are in general unique for a particular power-law segment in a given dynamical phase of the outflow. This also holds for all the $g$-factors appearing in eq. (\ref{C_pol}). Their values, however, are determined by matching them to numerical results (for a variety of $(p,k)$ values) and solving the resulting systems of equations. They are in fact the calibration of the flux-scalings.

\subsection{Flux-scalings}

\subsubsection{Flux-scalings during the BM phase}

A very similar approach has been taken by \citet{vanEerten2009}. These authors have explored optically thin synchrotron radiation from relativistic outflows, taking into account all the possible spectra that result from either fast or slow cooling (\citealt{Sari1998}). Here we expand on that by including self-absorption. Table \ref{F-rel-q} contains the values of the $q$-exponents (analytically derived dependencies), while Table \ref{F-rel-g} contains the values of the $g$-factors (numerically determined calibration).

The values of the $q$-exponents are in agreement with the formulas presented in \citet{Granot2002}, apart from the fact that we have chosen to include an extra parameter $\xi$. The normalisation of the flux-scalings results in slightly lower fluxes (of order $30\%$) compared to \citet{Granot2002} a difference which can be attributed to the varying adiabatic index of the simulations (\citealt{vanEerten2010a}). Below $\nu_{\textrm{m}}$ we have ignored stimulated emission associated with a population inversion of the electron distribution at the low-energy limit, something which \citet{Granot2002} have included in their model. That accounts for a factor of approximately $\left[3(p+2)/4\right]$ difference in flux between those predictions and the present ones.

\subsubsection{Flux-scalings during the ST phase}

For the Newtonian phase of the outflow, we repeat the same procedure as in the BM phase. The analytically derived dependencies are presented in Table \ref{F-nonrel-q} while the factors of the calibrating polynomials are presented in Table \ref{F-nonrel-g}. We note that the values of $q_{\nu}$ and $q_t$ are in agreement with those presented in \citet{Frail2000}, apart from their equation (A18) where the scaling for $\nu_{\textrm{obs}} \ll \nu_{\textrm{m}}$ is in error. This error also appears in \citet{vanEerten2010a}.

\begin{table}
\centering
\parbox{0.93\columnwidth}{\captionof{table}{$q$-exponents in the BM phase. Analytically derived $q$-exponents for self-absorbed and optically thin synchrotron radiation in the BM phase. The quantities on the left correspond to the different physical parameters on which the flux depends (see eq. \ref{F_nu}). Each column describes their values for a given power-law segment.\label{F-rel-q}}}
\small\addtolength{\tabcolsep}{+2pt}
\begin{tabular}{l c c c c} 
\toprule                        
 & $F_{\,2}$ & $F_{5\!/\!2}$ & $F_{1\!/\!3}$ & $F_{(1\!-\!p)\!/\!2}$ \\ [0.5ex] 
\hline                  
$q_{\xi}$ & $-1$ & $0$ & $\frac{5}{3}$ & $2-p$ \\ [1.5ex]
$q_{e}$ & $1$ & $0$ & $-\frac{2}{3}$ & $p-1$ \\ [1.5ex]
$q_{B}$ & $0$ & $-\frac{1}{4}$ & $\frac{1}{3}$ & $\frac{p+1}{4}$ \\ [1.5ex]
$q_n$ & $-\frac{2}{4-k}$ & $-\frac{2}{4-k}$ & $\frac{2}{4-k}$ & $\frac{2}{4-k}$ \\ [1.5ex]
$q_E$ & $\frac{2}{4-k}$ & $\frac{4+k}{4(4-k)}$ & $\frac{10-4k}{3(4-k)}$ & $\frac{12+4p-kp-5k}{4(4-k)}$ \\ [1.5ex]
$q_{\nu}$ & 2 &  $\frac{5}{2}$ & $\frac{1}{3}$ & $\frac{1-p}{2}$ \\ [1.5ex]
$q_t$ & $\frac{2}{4-k}$ & $\frac{20-3k}{4(4-k)}$ & $\frac{2-k}{4-k}$ & $\frac{12+3kp-5k-12p}{4(4-k)}$ \\ [1.5ex]
$q_z$ & $\frac{10-3k}{4-k}$ & $\frac{36-11k}{4(4-k)}$ & $\frac{10-k}{3(4-k)}$ & $\frac{12+4p-k-kp}{4(4-k)}$ \\ [1ex]      
\bottomrule       
\end{tabular}\par
\smallskip
\end{table}

\begin{table}
\centering
\parbox{0.91\columnwidth}{\captionof{table}{$g$-factors in the BM phase. Numerically determined $g$-factors for self-absorbed and optically thin synchrotron radiation in the BM phase. The quantities on the left correspond to different factors of the normalizing polynomial (see eq. \ref{C_pol}). Each column describes their values for a given power-law segment.\label{F-rel-g}}}
\small\addtolength{\tabcolsep}{+3pt}
\begin{tabular}{l c c c c}
\toprule
 & $F_{\,2}$ & $F_{5\!/\!2}$ & $F_{1\!/\!3}$ & $F_{(1\!-\!p)\!/\!2}$ \\ [0.5ex]
\hline
$g_0$ & $-18.350$ & $-26.170$ & $-3.232$ & $-6.689$ \\ [1.5ex]
$g_p$ & $0$ & $0$ & $0$ & $7.810$ \\ [1.5ex]
$g_{pp}$ & $0$ & $0$ & $0$ & $0.075$ \\ [1.5ex]
$g_k$ & $0.237$ & $0.185$ & $-0.262$ & $-0.286$ \\ [1.5ex]
$g_{kk}$ & $0.133$ & $0.120$ & $-0.014$ & $0.020$ \\ [1ex]
\bottomrule
\end{tabular}
\smallskip
\end{table}

\begin{table}
\centering
\parbox{0.98\columnwidth}{\captionof{table}{$q$-exponents in the ST phase. Analytically derived $q$-exponents for self-absorbed and optically thin synchrotron radiation in the ST phase. The quantities on the left correspond to the different physical parameters on which the flux depends (see eq. \ref{F_nu}). Each column describes their values for a given power-law segment.\label{F-nonrel-q}}}
\small\addtolength{\tabcolsep}{+2pt}
\begin{tabular}{l c c c c}
\toprule
 & $F_{\,2}$ & $F_{5\!/\!2}$ & $F_{1\!/\!3}$ & $F_{\!(\!1\!-\!p\!)\!/\!2}$ \\ [0.5ex] 
\hline
$q_{\xi}$ & $-1$ & $0$ & $\frac{5}{3}$ & $2-p$ \\ [1.5ex]
$q_{e}$ & $1$ & $0$ & $-\frac{2}{3}$ & $p-1$ \\ [1.5ex]
$q_{B}$ & $0$ & $-\frac{1}{4}$ & $\frac{1}{3}$ & $\frac{p+1}{4}$ \\ [1.5ex]
$q_n$ & $-\frac{4}{5-k}$ & $-\frac{11}{4(5-k)}$ & $\frac{13}{3(5-k)}$ & $\frac{19-5p}{4(5-k)}$ \\ [1.5ex]
$q_E$ & $\frac{4}{5-k}$ & $\frac{6+k}{4(5-k)}$ & $\frac{7-4k}{3(5-k)}$ & $\frac{10p+6-k(p+5)}{4(5-k)}$ \\ [1.5ex]
$q_{\nu}$ & $2$ & $\frac{5}{2}$ & $\frac{1}{3}$ & $\frac{1-p}{2}$ \\ [1.5ex]
$q_t$ & $\frac{2k-2}{5-k}$ & $\frac{11}{2(5-k)}$ & $\frac{24-10k}{3(5-k)}$ & $\frac{3(7-5p)+4k(p-2)}{2(5-k)}$ \\ [1.5ex]
$q_z$ & $\frac{17-5k}{5-k}$ & $\frac{24-7k}{2(5-k)}$ & $\frac{6k-4}{3(5-k)}$ & $\frac{5(2p+k)-3(2+kp)}{2(5-k)}$ \\ [1ex]
\bottomrule 
\end{tabular}
\smallskip
\end{table}

\begin{table}
\centering  
\parbox{0.9\columnwidth}{\captionof{table}{$g$-factors in the ST phase. Numerically determined $g$-factors for self-absorbed and optically thin synchrotron radiation in the ST phase. The quantities on the left correspond to different factors of the normalizing polynomial (see eq. \ref{C_pol}). Each column describes their values for a given power-law segment.\label{F-nonrel-g}}}
\small\addtolength{\tabcolsep}{+3pt}
\begin{tabular}{l c c c c c} 
\toprule
 & $F_{\,2}$ & $F_{5\!/\!2}$ & $F_{1\!/\!3}$ & $F_{\!(\!1\!-\!p\!)\!/\!2}$ \\ [0.5ex] 
\hline
$g_0$ & $-16.510$ & $-25.645$ & $-5.513$ & $-8.789$ \\ [1.5ex]
$g_p$ & $0$ & $0$ & $0$ & $8.528$ \\ [1.5ex]
$g_{pp}$ & $0$ & $0$ & $0$ & $0.230$ \\ [1.5ex]
$g_k$ & $0.126$ & $-0.017$ & $-0.044$ & $0.157$ \\ [1.5ex]
$g_{kk}$ & $-0.009$ & $-0.014$ & $0.008$ & $0.070$ \\ [1ex]
\bottomrule 
\end{tabular}
\smallskip
\end{table}

\subsection{The sharpness of spectral breaks}

In practice the spectral breaks are not infinitely sharp as shown in Fig. \ref{fig_spec_am} and \ref{fig_spec_ma} but show a gradual transition from one power-law index to another. To complete our description of instantaneous spectra, we need to provide a formula for the sharpness of spectral breaks.

An approach commonly used \citep{Granot2002,vanEerten2009} is to describe the flux close to a break by the following equation (\citealt{Beuermann1999})

\begin{equation}\label{break_nu}
F_{\nu}(\nu_{\textrm{obs}})= A\, \left[ \left( \frac{\nu_{\textrm{obs}}}{\nu_0} \right)^{-a_1 \, s} + \left( \frac{\nu_{\textrm{obs}}}{\nu_0} \right)^{-a_2 \, s} \right] ^{-1/s},
\end{equation}
where ($\nu_0,A$) are the coordinates of the meeting point of the two power laws associated with the break, $a_1$ and $a_2$ are the asymptotic power-law indices before and after the break, respectively, and $s$ is the so called `sharpness parameter'. We have performed $\chi ^2 \!$-minimization fitting in logarithmic space to obtain values of $s$ for specific runs and used those to arrive at a description of the sharpness in terms of a polynomial of $p$ and $k$. This polynomial has the general form

\begin{equation}\label{s(p,k)}
s = s_0 + s_p\, p + s_k\, k + s_{kk}\, k^2.
\end{equation}
Its factors have been determined by solving the system of equations resulting from the application of eq. (\ref{break_nu}) to models with different $p$ and $k$ parameters. In Tables \ref{s-rel} and \ref{s-nonrel} we present the values of $s_0, s_p, s_k$ and $s_{kk}$ in the BM and the ST phase, respectively.

\begin{figure}
\includegraphics[width=\columnwidth]{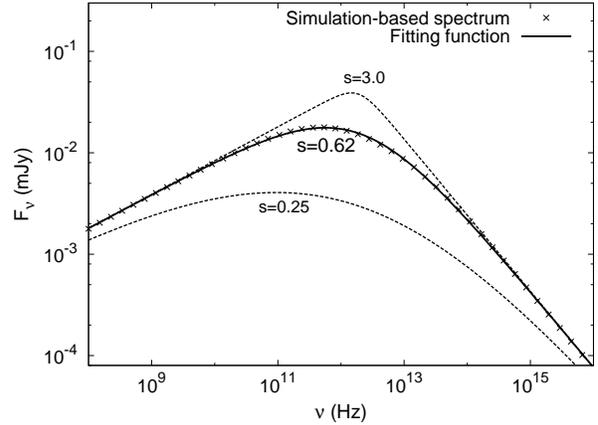}
\caption{The effect of sharpness on the flux close to a spectral break. This fragment of the simulation-based spectrum focuses on the flux around $\nu_{\textrm{m}}$. The best fit is shown along with two more curves that have the same parameters but different sharpness. Simulation-based spectrum has the following model parameters: $E_{52}=1,\, n_0=1,\, p=2.5,\, k=0,\, \xi=10^{-2},\, \epsilon_{\textrm{e}}=10^{-1},\, \epsilon_{\textrm{B}}=10^{-2},\, d_{28}=1,\, z=0.56,\, t_{\textrm{obs}}=100\,\textrm{days}$.}
\label{fig_spec_break}
\end{figure}

In Fig. \ref{fig_spec_break} a best fit to the shape of the spectrum around $\nu_{\textrm{m}}$ is shown, for one of the run models. We also plot the flux for two different values of $s$ to illustrate the notable effect it can have on flux levels.

\begin{table}
\centering  
\parbox{0.9\columnwidth}{\captionof{table}{$s$-factors in the BM phase. Numerically determined $s$-factors (see eq. \ref{s(p,k)}) for all possible breaks in the BM phase. Each column describes a specific break, with the two associated spectral indices denoted on top.\label{s-rel}}}
\small\addtolength{\tabcolsep}{+2pt}
\begin{tabular}{l c c c c} 
\toprule
 & $2\! \rightarrow \! \dfrac{5}{2}$ & $\dfrac{5}{2}\! \rightarrow \! \dfrac{1-p}{2}$ & $2\! \rightarrow \! \dfrac{1}{3}$ & $\dfrac{1}{3} \! \rightarrow \! \dfrac{1-p}{2}$ \\ [0.5ex] 
\hline
$s_0$ & $-2.91$ & $1.24$ & $1.64$ & $1.83$ \\ [1.5ex]
$s_p$ & $-0.11$ & $-0.145$ & $0$ &-0.41$$ \\ [1.5ex]
$s_k$ & $0.04$ & $0$ & $-0.18$ & $0$ \\ [1.5ex]
$s_{kk}$ & $0$ & $0$ & $0$ & $0$ \\ [1ex]
\bottomrule 
\end{tabular}
\smallskip 
\end{table}

\begin{table}
\centering  
\parbox{0.9\columnwidth}{\captionof{table}{$s$-factors in the ST phase. Numerically determined $s$-factors (see eq. \ref{s(p,k)}) for all possible breaks in the ST phase. Each column describes a specific break, with the two associated spectral indices denoted on top.\label{s-nonrel}}}
\small\addtolength{\tabcolsep}{+2pt}
\begin{tabular}{l c c c c} 
\toprule
 & $2\! \rightarrow \! \dfrac{5}{2}$ & $\dfrac{5}{2}\! \rightarrow \! \dfrac{1-p}{2}$ & $2\! \rightarrow \! \dfrac{1}{3}$ & $\dfrac{1}{3} \! \rightarrow \! \dfrac{1-p}{2}$ \\ [0.5ex] 
\hline
$s_0$ & $-5.50$ & $3.50$ & $2.63$ & $1.88$ \\ [1.5ex]
$s_p$ & $0.73$ & $-0.71$ & $-0.24$ & $-0.46$ \\ [1.5ex]
$s_k$ & $0.10$ & $-0.07$ & $-0.31$ & $0.11$ \\ [1.5ex]
$s_{kk}$ & $0$ & $-0.11$ & $-0.07$ & $-0.02$ \\ [1ex]
\bottomrule 
\end{tabular}
\smallskip
\end{table}

\pagebreak


\section{The transrelativistic regime}\label{transrel}
The results of the previous section constitute a full description of the possible synchrotron spectra (ignoring cooling) during the ultrarelativistic and non-relativistic dynamical phases of the afterglow evolution. However, we have not yet addressed a large portion of the afterglow's overall behaviour, namely the \textit{transrelativistic} regime. During this stage the dynamics deviates considerably from the BM solution, without having settled yet into the ST solution. As mentioned in section \ref{intro}, this phase of the afterglow typically spans a few orders of magnitude in observer time, while there is no full description of its dynamics, even in the simple, spherical case.

An approach we have investigated and found useful is that of treating the transrelativistic phase as a `break' during which the temporal evolution of the spectrum's critical parameters displays a smooth transition from the relativistic power-law behaviour to the non-relativistic one. These parameters could be, for example, the values of the flux at every possible power-law segment. However, the same level of accuracy can be achieved by using the positions of the critical frequencies and the flux at one of them, instead. Based on values for these parameters one can construct the spectrum \citep{Sari1998, Wijers1999} because the slopes of the power-law segments are known for a given ordering of $\nu_{\textrm{a}}$ and $\nu_{\textrm{m}}$.

\subsection{Peak flux}

A convenient frequency to measure the flux is at $\nu_{\textrm{m}}$ of spectrum 1. This is because we can assume that the bulk of the electrons radiate most of their power at that frequency. Using eq. (\ref{F-thin}) in combination with the scalings for the dynamics in each of the two extreme phases of the outflow, we find for the flux at $\nu_{\textrm{m}}$

\begin{equation}\label{FmBM}
F_{\textrm{\!m-BM}} = C_{\textrm{pol}} \, \xi \, \epsilon_{\textrm{B}}^{\frac{1}{2}} \, n_0^{\frac{4}{2(4-k)}} \, E_{52}^{\frac{8-3k}{2(4-k)}} \, t_{\textrm{obs}}^{\frac{-k}{2(4-k)}} (1\!+\!z)^{\frac{8-k}{2(4-k)}}\, d_{28}^{-2} ,
\end{equation}
in the BM phase and
\begin{equation}\label{FmST}
F_{\textrm{\!m-ST}} = C_{\textrm{pol}} \, \xi \, \epsilon_{\textrm{B}}^{\frac{1}{2}} \, n_0^{\frac{7}{2(5-k)}} \, E_{52}^{\frac{8-3k}{2(5-k)}} \, t_{\textrm{obs}}^{\frac{3-2k}{5-k}} (1\!+\!z)^{\frac{2+k}{5-k}} \, d_{28}^{-2} ,
\end{equation}
in the ST phase. The $g$-factors of $C_{\textrm{pol}}$ for both dynamical phases are presented in Table \ref{F_m-table}.

\begin{table}
\centering  
\parbox{0.53\columnwidth}{\captionof{table}{$g$-factors for $F_{\textrm{\!m-BM}}$ and $F_{\textrm{\!m-ST}}$. Numerically determined $g$-factors for $F_{\textrm{\!m}}$ (measured at $\nu_{\textrm{m}1}$), both in the BM and ST phase.\label{F_m-table}}}
\small\addtolength{\tabcolsep}{+4pt}
\begin{tabular}{l c c} 
\toprule
 & $F_{\textrm{\!m-BM}}$ & $F_{\textrm{\!m-ST}}$ \\ [0.5ex] 
\hline
$g_0$ & $0.531$ & $-0.674$ \\ [1.5ex]
$g_p$ & $0.487$ & $0.305$ \\ [1.5ex]
$g_{pp}$ & $-0.060$ & $-0.019$ \\ [1.5ex]
$g_k$ & $-0.291$ & $-0.055$ \\ [1.5ex]
$g_{kk}$ & $0.004$ & $0.015$ \\ [1ex]
\bottomrule
\end{tabular}
\smallskip
\end{table}

\subsection{Critical frequencies}

The behaviour of the critical frequencies can easily be deduced (in either the BM or ST phase) by equating the flux formulas on both sides of a spectral break. For each of $\nu_{\textrm{m}}$ and $\nu_{\textrm{a}}$ there will be two such expressions corresponding to the two possible spectra for the two different orderings of the frequencies. In general, the value of a critical frequency will be given by the following formula

\begin{equation}\label{nu_cr}
\nu_{\textrm{cr}} = f_{\textrm{n}}\, \xi^{q_{\xi}} \, \epsilon_{\textrm{e}}^{q_e} \, \epsilon_{\textrm{B}}^{q_B} \, n_0^{q_n} \, E_{52}^{q_E} \, \nu_{\textrm{obs}}^{q_{\nu}} \, t_{\textrm{obs}}^{q_t} \, (1+z)^{q_z}.
\end{equation}
The numerical factors $f_{\textrm{n}}$ result from equating the fluxes of the power-laws at each side of the spectral break. Tables \ref{nu_cr-BM-am} and \ref{nu_cr-BM-ma} summarize the formulas for the critical frequencies in the BM phase for the two different possible spectra. For the ST phase we repeat the same procedure and summarize our results in Tables \ref{nu_cr-ST-am} and \ref{nu_cr-ST-ma}.

For clearer presentation we have labelled as $\nu_{\textrm{a}1}$ and $\nu_{\textrm{m}1}$ the critical frequencies $\nu_{\textrm{a}}$ and $\nu_{\textrm{m}}$, respectively, when $\nu_{\textrm{a}} < \nu_{\textrm{m}}$ (i.e. when spectrum 1 applies), while they are labelled as $\nu_{\textrm{a}2}$ and $\nu_{\textrm{m}2}$ in the opposite case when spectrum 2 applies.

\begin{table}
\centering
\parbox{0.98\columnwidth}{\captionof{table}{$f_{\textrm{n}}$ and $q$-exponents for critical frequencies in the BM phase, while $\nu_{\textrm{a}}<\nu_{\textrm{m}}$ (spectrum 1). The $q$-exponents carry the analytically derived dependencies, while the $f_{\textrm{n}}$-factors carry the flux-calibrating $C_{\textrm{pol}}$ and $h(p)$ (see eq. \ref{F_nu}, \ref{C_pol} and \ref{h(p)}).\label{nu_cr-BM-am}}}
\small\addtolength{\tabcolsep}{+10pt}
\begin{tabular}{l c c}
\toprule
 & $\nu_{\textrm{a}1}$ & $\nu_{\textrm{m}1}$ \\ [0.5ex] 
\hline
$f_{\textrm{n}}$ & $\left(\frac{C_{\!1\!/\!3} \, h_{\!1\!/\!3} }{C_{\!2} \, h_{\!2} }\right)^{\frac{3}{5}}$ & $\left(\frac{C_{\!\!(1\!-\!p)\!/\!2} \, h_{\!(1\!-\!p)\!/\!2} }{C_{\!1\!/\!3} \, h_{\!1\!/\!3}}\right)^{\frac{6}{3p-1}}$ \\ [2.5ex]
$q_{\xi}$ & $\frac{8}{5}$ & $-2$ \\ [1.5ex]
$q_{e}$ & $-1$ & $2$ \\ [1.5ex]
$q_{B}$ & $\frac{1}{5}$ & $\frac{1}{2}$ \\ [1.5ex]
$q_n$ & $\frac{12}{5(4-k)}$ & $0$ \\ [1.5ex]
$q_E$ & $\frac{4-4k}{5(4-k)}$ & $\frac{1}{2}$ \\ [1.5ex]
$q_t$ & $-\frac{3k}{5(4-k)}$ & $-\frac{3}{2}$ \\ [1.5ex]
$q_z$ & $\frac{8k-20}{5(4-k)}$ & $\frac{1}{2}$ \\ [1ex]
\bottomrule 
\end{tabular}
\smallskip
\end{table}

\begin{table}
\centering
\parbox{0.97\columnwidth}{\captionof{table}{$f_{\textrm{n}}$ and $q$-exponents for critical frequencies in the BM phase, while $\nu_{\textrm{m}}<\nu_{\textrm{a}}$ (spectrum 2).\label{nu_cr-BM-ma}}}
\small\addtolength{\tabcolsep}{+10pt}
\begin{tabular}{l c c}
\toprule
 & $\nu_{\textrm{m}2}$ & $\nu_{\textrm{a}2}$ \\ [0.5ex] 
\hline
$f_{\textrm{n}}$ & $\left(\frac{C_{\!2}\,h_{\!2}}{C_{\!5\!/\!2}\,h_{\!5\!/\!2}}\right)^2$ & $\left(\frac{C_{\!\!(1\!-\!p)\!/\!2} \,h_{\!(1\!-\!p)\!/\!2} } {C_{\!5\!/\!2} \, h_{\!5\!/\!2}}\right)^{\frac{2}{4+p}}$ \\ [2.5ex]
$q_{\xi}$ & $-2$ & $\frac{4-2p}{4+p}$ \\ [1.5ex]
$q_{e}$ & $2$ & $\frac{2(p-1)}{4+p}$ \\ [1.5ex]
$q_{B}$ & $\frac{1}{2}$ & $\frac{p+2}{2(4+p)}$ \\ [1.5ex]
$q_n$ & $0$ & $\frac{8}{(4+p)(4-k)}$ \\ [1.5ex]
$q_E$ & $\frac{1}{2}$ & $\frac{8+4p-kp-6k}{2(4+p)(4-k)}$ \\ [1.5ex]
$q_t$ & $-\frac{3}{2}$ & $\frac{3kp-2k-12p-8}{2(4+p)(4-k)}$ \\ [1.5ex]
$q_z$ & $\frac{1}{2}$ & $\frac{10k+4p-24-kp}{2(4+p)(4-k)}$ \\ [1ex]
\bottomrule 
\end{tabular}
\smallskip
\end{table}

\begin{table}
\centering
\parbox{0.98\columnwidth}{\captionof{table}{$f_{\textrm{n}}$ and $q$-exponents for critical frequencies in the ST phase, while $\nu_{\textrm{a}}<\nu_{\textrm{m}}$ (spectrum 1). The expressions contained in $f_{\textrm{n}}$ need to be evaluated using the formulas applicable to the ST regime.\label{nu_cr-ST-am}}}
\small\addtolength{\tabcolsep}{+10pt}
\begin{tabular}{l c c}
\toprule
 & $\nu_{\textrm{a}1}$ & $\nu_{\textrm{m}1}$ \\ [0.5ex] 
\hline
$f_{\textrm{n}}$ & $\left(\frac{C_{\!1\!/\!3} \, h_{\!1\!/\!3} }{C_{\!2} \, h_{\!2} }\right)^{\frac{3}{5}}$ & $\left(\frac{C_{\!\!(1\!-\!p)\!/\!2} \, h_{\!(1\!-\!p)\!/\!2} }{C_{\!1\!/\!3} \, h_{\!1\!/\!3}}\right)^{\frac{6}{3p-1}}$ \\ [2.5ex]
$q_{\xi}$ & $\frac{8}{5}$ & $-2$ \\ [1.5ex]
$q_{e}$ & $-1$ & $2$ \\ [1.5ex]
$q_{B}$ & $\frac{1}{5}$ & $\frac{1}{2}$ \\ [1.5ex]
$q_n$ & $\frac{5}{5-k}$ & $-\frac{5}{2(5-k)}$ \\ [1.5ex]
$q_E$ & $-\frac{5+4k}{5(5-k)}$ & $\frac{10-k}{2(5-k)}$ \\ [1.5ex]
$q_t$ & $\frac{30-16k}{5(5-k)}$ & $\frac{4k-15}{5-k}$ \\ [1.5ex]
$q_z$ & $\frac{21k-55}{5(5-k)}$ & $\frac{10-3k}{5-k}$ \\ [1ex]
\bottomrule 
\end{tabular}
\smallskip
\end{table}

\begin{table}
\centering
\parbox{0.97\columnwidth}{\captionof{table}{$f_{\textrm{n}}$ and $q$-exponents for critical frequencies in the ST phase, while $\nu_{\textrm{m}}<\nu_{\textrm{a}}$ (spectrum 2).\label{nu_cr-ST-ma}}}
\small\addtolength{\tabcolsep}{+10pt}
\begin{tabular}{l c c}
\toprule
 & $\nu_{\textrm{m}2}$ & $\nu_{\textrm{a}2}$ \\ [0.5ex] 
\hline
$f_{\textrm{n}}$ & $\left(\frac{C_{\!2}\,h_{\!2}}{C_{\!5\!/\!2}\,h_{\!5\!/\!2}}\right)^2$ & $\left(\frac{C_{\!\!(1\!-\!p)\!/\!2} \,h_{\!(1\!-\!p)\!/\!2} } {C_{\!5\!/\!2} \, h_{\!5\!/\!2}}\right)^{\frac{2}{4+p}}$ \\ [2.5ex]
$q_{\xi}$ & $-2$ & $\frac{4-2p}{4+p}$ \\ [1.5ex]
$q_{e}$ & $2$ & $\frac{2(p-1)}{4+p}$ \\ [1.5ex]
$q_{B}$ & $\frac{1}{2}$ & $\frac{p+2}{2(4+p)}$ \\ [1.5ex]
$q_n$ & $-\frac{5}{2(5-k)}$ & $\frac{30-5p}{2(4+p)(5-k)}$ \\ [1.5ex]
$q_E$ & $\frac{10-k}{2(5-k)}$ & $\frac{10p-kp-6k}{2(4+p)(5-k)}$ \\ [1.5ex]
$q_t$ & $\frac{4k-15}{5-k}$ & $\frac{10-8k-15p+4kp}{(4+p)(5-k)}$ \\ [1.5ex]
$q_z$ & $\frac{10-3k}{5-k}$ & $\frac{12k+10p-30-3kp}{(4+p)(5-k)}$ \\ [1ex]
\bottomrule 
\end{tabular}
\smallskip
\end{table}

\subsection{Evolution of critical parameters}\label{evo-cr-par}

A practical way of describing the temporal evolution of the parameters needed to construct a spectrum at any point is that of a smoothly broken power law. We can use eq. (\ref{break_nu}), this time characterizing a temporal break in the following manner

\begin{equation}\label{break_t_ini}
\Phi(t_{\textrm{obs}})= A\, \left[ \left( \frac{t_{\textrm{obs}}}{t_0} \right)^{-a_1 \, s_{\textrm{t}}} + \left( \frac{t_{\textrm{obs}}}{t_0} \right)^{-a_2 \, s_{\textrm{t}}} \right] ^{-1/s_{\textrm{t}}},
\end{equation}
where $(t_0,A)$ is the meeting point of the asymptotes and $\Phi$ is the value of any of the critical parameters. In this version of eq. (\ref{break_nu}) $a_1$ and $a_2$ are the BM and ST slopes, respectively. We can rewrite the above equation in the following way

\begin{equation}\label{break_t}
\Phi(t_{\textrm{obs}})= \left( \, \Phi_{\textrm{BM}}^{- s_{\textrm{t}}} + \Phi_{\textrm{ST}}^{- s_{\textrm{t}}} \, \right) ^{-1/s_{\textrm{t}}},
\end{equation}
where both $\Phi_{\textrm{BM}}$ and $\Phi_{\textrm{ST}}$ have to be evaluated at $t_{\textrm{obs}}$.

In the previous sections we have established not only the scalings of the critical parameters we wish to follow, but their actual values as a function of observer time in both extreme dynamical regimes. This allows us to insert them directly into eq. (\ref{break_t}), where the only unknown left is the sharpness $s_{\textrm{t}}$. As in the case of spectral breaks we have performed $\chi ^2 \!$-minimization fitting in logarithmic space and have arrived at a description of `$s_{\textrm{t}}$' in terms of a polynomial of the following form

\begin{equation}\label{s-crit}
s_{\textrm{t}} = s_0 + s_p\, p + s_{pp}\,p^2 + s_k\, k + s_{kk}\, k^2.
\end{equation}
Results for the values of these $s_{\textrm{t}}$-factors are presented in Table \ref{s-critical}.

\begin{figure}
\includegraphics[width=\columnwidth]{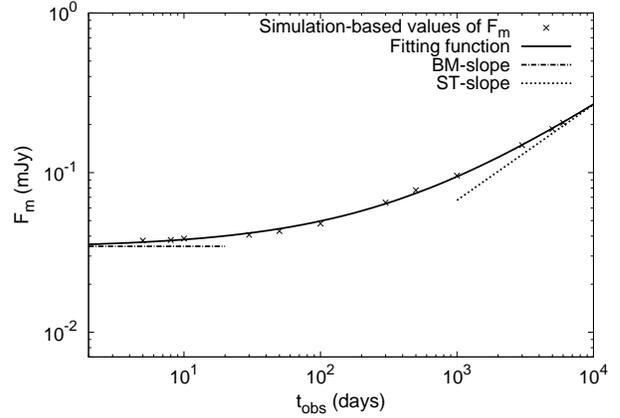}
\caption{A broken power-law fit to the evolution of $F_{\textrm{m}}$. Plotted are the BM and ST asymptotes. Simulation-based data have the following model parameters: $E_{52}=1,\, n_0=1,\, p=2.5,\, k=0,\, \xi=10^{-2},\, \epsilon_{\textrm{e}}=10^{-1},\, \epsilon_{\textrm{B}}=10^{-2},\, d_{28}=1,\, z=0.56$.}
\label{fig_F_m_fit}
\end{figure}

An example fit of $F_{\textrm{m}}$ is shown in Fig. \ref{fig_F_m_fit}. The behaviour of $F_{\textrm{m}}$ displays clear deviations from the BM scalings already before $100$ days observer time, for $E_{\textrm{iso}}=10^{52}\, \textrm{erg}$, $n_0=1$ and $k=0$. It settles to values sufficiently close (within $10\%$) to the ST solution at around $5000$ days. The \textit{duration} of the transrelativistic regime is represented in $s_{\textrm{t}}$, for a given set of physical parameters. Table \ref{s-critical} demonstrates that the sharpness of every parameter is generally unique. Based on that we conclude that duration and features of the transrelativistic phase will be manifested differently across the spectrum.

\begin{table}
\centering  
\parbox{0.92\columnwidth}{\captionof{table}{$s_{\textrm{t}}$-factors for the evolution of critical parameters. Numerically determined $s_{\textrm{t}}$-factors (see eq. \ref{s-crit}) describing the evolution of critical parameters (break-frequencies and maximum flux) from the BM to the ST phase. Each column describes a specific parameter, the name of which is denoted on top.\label{s-critical}}}
\small\addtolength{\tabcolsep}{+2pt}
\begin{tabular}{l c c c c c} 
\toprule
 & $F_{\textrm{m}}$ & $\nu_{\textrm{a}1}$ & $\nu_{\textrm{a}2}\, ^{a}$ & $\nu_{\textrm{m}1}$ & $\nu_{\textrm{m}2}$ \\ [0.5ex] 
\hline
$s_0$ & $-1.49$ & $-0.61$ & $22.50$ & $-0.89$ & $0.43$ \\ [1.5ex]
$s_p$ & $0.09$ & $0$ & $-5.00$ & $1.12$ & $0$ \\ [1.5ex]
$s_{pp}$ & $0$ & $0$ & $0$ & $-0.21$ & $0$ \\ [1.5ex]
$s_k$ & $-0.76$ & $-0.12$ & $-2.00$ & $0.14$ & $0$ \\ [1.5ex]
$s_{kk}$ & $0.12$ & $-0.02$ & $0$ & $0$ & $0$ \\ [1ex]
\bottomrule\\ [0.1ex]
\multicolumn{6}{p{0.4\textwidth}}{$^{a}$The $s$-factors for $\nu_{\textrm{a}2}$ have not been determined by solving the system of equations resulting from measuring its value in different models but comprise a rather heuristic approach that minimizes the deviations from the numerically determined values.}
\end{tabular}
\smallskip
\end{table}

\subsection{Evolution of the sharpness of spectral breaks}

Our findings so far enable us to determine the values of the critical frequencies and $F_{\textrm{m}}$ at any given time for a burst of given physical properties. This allows for an accurate calculation of the flux at any given power-law segment of the spectrum. What is left to specify is the flux close to a spectral break for a general $t_{\textrm{obs}}$. To achieve that we need to provide a quantitative description for the evolution of the sharpness of spectral breaks from the BM to the ST phase.

\begin{figure}
\includegraphics[width=\columnwidth]{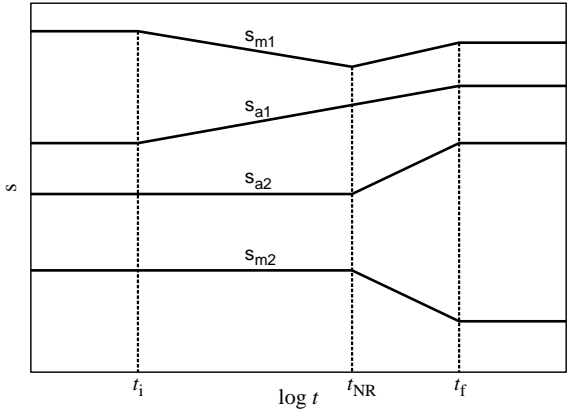}
\caption{Evolution of sharpness for all possible spectral breaks. Before $t_{\textrm{i}}$ and after $t_{\textrm{f}}$, $s$ resumes its standard BM and ST values. The sharpness of the break around $\nu_{\textrm{m}1}$ shows the most complex pattern, by declining initially during the transrelativistic regime and then rising again to meet its ST asymptote. The value of $s_{\textrm{m}1}$ at $t_{\textrm{NR}}$ is $\sim 0.18$ smaller than the BM value.}
\label{evo-s}
\end{figure}

As it turns out the sharpness of every spectral break follows a `characteristic path' as it evolves from the relativistic values to the Newtonian ones. This path is qualitatively independent of the physical properties of the burst and is unique for every spectral break. In Fig. \ref{evo-s} we present these paths for all possible breaks. From the data gathered we have identified three timescales that are represented in this figure: $t_{\textrm{i}}$, $t_{\textrm{NR}}$ and $t_{\textrm{f}}$. In fact, we can simplify things further by setting $t_{\textrm{i}}=t_{\textrm{NR}}/100$ and $t_{\textrm{f}}=10 \, t_{\textrm{NR}}$ which is generally valid up to a few percent, regardless of the physical parameters of a burst.

Determination of $t_{\textrm{NR}}$ carries (as in the case of fluxes and critical frequencies) an analytic and a numerical component. The analytic part is motivated by considerations of the dynamics and is similar to other estimates of an observer time marking the transition to the Newtonian phase \citep{Livio2000,Piran2004}. Specifically it is identified as the observer time at which the shock Lorentz factor drops to the value of 2, following the BM solution.

\begin{equation}
t_{\textrm{NR}} \sim \frac{8^{\frac{k-4}{3-k}}}{4-k} \, A_{\textrm{NR}}^{\frac{1}{3-k}} \, (1+z),
\end{equation}
where $A_{\textrm{NR}} = \left[ \left( \dfrac{17}{8\, \pi \, m_{\textrm{p}}} \right) \, c^{k-5}\, E_{52} \, n_0^{-1} \right] $.

\vspace{0.3cm}

Including the numerical calibration the expression for $t_{\textrm{NR}}$ takes the form

\begin{equation}\label{tnr}
t_{\textrm{NR}} = 10^{13.66} \, \, \frac{8^{\frac{k-4}{3-k}}}{4-k} \, A_{\textrm{NR}}^{\frac{1}{3-k}} \, (1+z) \, \, \textrm{days}.
\end{equation}

Like all other timescales describing a transition from BM to ST, $t_{\textrm{NR}}$ scales as $(E_{52}/n_0)^{1/(3-k)}\,(1+z)$, as we expect from dimensional analysis (\citealt{vanEerten2012a}). The same holds for $t_0$ appearing in eq. (\ref{break_t_ini}) for all critical spectrum parameters. However, the actual value of $t_0$ for every parameter is influenced by the flux-calibrating polynomials $C_{\!\textrm{pol}}$ (eq. \ref{C_pol}) and is therefore in general unique.

The equations, tables and plots of sections \ref{flux-prescriptions} and \ref{transrel} carry all the information necessary to construct a spectrum at any given time, based on given values for the relevant physical parameters that we have discussed in this work. In the following section we demonstrate how these results can be used to calculate the observed flux at any given frequency and time.


\section{Using the prescriptions}\label{using}

In this section we focus on the practical side of this work which is to construct spectra at any given observer time based on values for the physical parameters of a burst. These parameters we repeat here for clarity: $E_{52}$ (isotropic blast-wave energy in units of $10^{52}\,\textrm{erg}$), $n_0$ (number density at $10^{17}\,\textrm{cm}$), $p$ (index of the electron power-law distribution), $k$ (index of the density distribution of the matter surrounding the burster), $\xi$ (fraction of accelerated electrons), $\epsilon_{\textrm{e}}$ and $\epsilon_{\textrm{B}}$ (fractions of internal energy assigned to the relativistic electrons and magnetic field, respectively), $\nu_{\textrm{obs}}$ and $t_{\textrm{obs}}$ (frequency and time of observation), $d_{28}$ and $z$ (luminosity distance in units of $10^{28}\,\textrm{cm}$ and redshift).

The task of constructing a spectrum out of the presented formulas can be divided in four parts. The first is to obtain the values of the critical frequencies and the maximum flux at a given observer time. The next step is to determine the shape of the spectrum and its general characteristics, i.e. values of the flux away from the breaks, for each of the two possible spectra. The third step is to assign the appropriate sharpness parameters to all spectral breaks. The fourth and final step is to use eq. (\ref{eq-spectrum}) and (\ref{final_flux}) in order to calculate the observed flux at a given frequency.

Each of the two spectra is described by a single equation, at a given observer time. This equation should essentially represent a mathematical formulation of a double-broken power law. One of the ways to achieve that (\citealt{Granot2002}) is to use a heuristic formula that combines eq. (\ref{break_nu}) with a factor that assigns a second break at a different frequency. Then the whole spectrum can be described by the following expression

\begin{eqnarray}\label{eq-spectrum}
F_{\nu}(\nu_{\textrm{obs}}) & = &  A\, \left[ \left( \frac{\nu_{\textrm{obs}}}{\nu_0} \right)^{-a_1 \, s} + \left( \frac{\nu_{\textrm{obs}}}{\nu_0} \right)^{-a_2 \, s} \right] ^{-1/s} \,\, \times \nonumber \\
 & & \left[ 1+\left( \frac{\nu_{\textrm{obs}}}{\nu_1} \right)^{h(a_2-a_3)} \right]^{-1/h}.
\end{eqnarray}
The first two terms on the right-hand side of the above equation describe the first break as usual, while the third term describes the second break. We have introduced $\nu_1$ the frequency of the second break, $a_3$ the slope of the third power-law segment and $h$ the sharpness of the second break.

Equation (\ref{eq-spectrum}) is exact only when the two breaks of the spectrum are sufficiently away from each other so that the power law connecting them is apparent, even for a small range of frequencies. When this is not the case, this equation provides an approximation to the real spectrum (\citealt{Granot2002}).

In Fig. \ref{flowchart} we present a flowchart of the basic steps towards creating a spectrum of the emitted radiation at any given observer time. The details of each step can be found in the following subsections of the text.

\tikzstyle{decision} = [diamond, draw, fill=blue!20, 
    text width=4.5em, text badly centered, node distance=2cm, inner sep=0pt]
\tikzstyle{block} = [rectangle, draw, fill=white!20, 
    text width=20em, text centered, rounded corners, minimum height=4em]
\tikzstyle{line} = [draw, -latex']
\tikzstyle{cloud} = [draw, ellipse,fill=red!20, node distance=3cm,
    minimum height=2em]

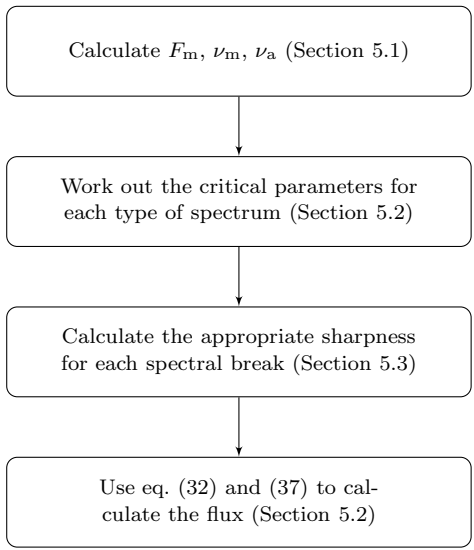
\begin{figure}
\begin{center} 
\begin{tikzpicture}[node distance = 2cm, auto]
    \node [block] (init) {Calculate $F_{\textrm{m}},\,\nu_{\textrm{m}},\,\nu_{\textrm{a}}$ (Section \ref{Fnu-values})};
    \node [block, below of=init] (decision) {Work out the critical parameters for each type of spectrum (Section \ref{Spectrum_shape})};
    \node [block, below of=decision] (evaluate) {Calculate the appropriate sharpness for each spectral break (Section \ref{sharp_spectral})};
    \node [block, below of=evaluate, node distance=2cm] (finish) {Use eq. (\ref{eq-spectrum}) and (\ref{final_flux}) to calculate the flux (Section \ref{Spectrum_shape})};
    \path [line] (init) -- (decision);
    \path [line] (decision) -- (evaluate);
    \path [line] (evaluate) -- (finish);
\end{tikzpicture}
\caption{Flowchart showing the basic steps of the presented method. References are given to specific sections of the paper where the steps are described in more detail.}\label{flowchart}
\end{center}
\end{figure}

\subsection{Values of $F_{\textrm{m}}$, $\nu_{\textrm{m}}$ and $\nu_{\textrm{a}}$}\label{Fnu-values}

In order to decide which of the two possible synchrotron spectra is valid (see Fig. \ref{fig_spec_am} and \ref{fig_spec_ma}) one needs to calculate all $\nu_{\textrm{a}1}$, $\nu_{\textrm{a}2}$, $\nu_{\textrm{m}1}$ and $\nu_{\textrm{m}2}$. While all these frequencies can in principle be calculated independently, we opt for a different method. Requiring that the flux at both ends of the spectrum (in the power-laws $\nu^{2}$ and $\nu^{1-p/2}$) be the same regardless of the type of spectrum, one can show that only three of the four frequencies are independent. The relation between them is

\begin{equation}\label{closure} 
\nu_{\textrm{a}2} = \nu_{\textrm{a}1}^{\frac{10}{3(4+p)}}\, \nu_{\textrm{m}1}^{\frac{3p-1}{3(4+p)}} \nu_{\textrm{m}2}^{\frac{1}{4+p}}.
\end{equation}

We have expressed $\nu_{\textrm{a}2}$ in terms of the others because its transrelativistic profile is the one deviating more from the broken power-law approach. This way inconsistencies that may arise during a spectral transition, due to the over-specification of the evolution of the spectrum, are avoided and the flux in the leftmost and rightmost power laws is independent of the ordering of the critical frequencies.

Equation (\ref{break_t}) should be used to calculate $\nu_{\textrm{a}1}$, $\nu_{\textrm{m}1}$ and $\nu_{\textrm{m}2}$ for a given set of physical parameters and for the same observer time, before applying eq. (\ref{closure}) to obtain $\nu_{\textrm{a}2}$. $\Phi_{\textrm{BM}}(t_{\textrm{obs}})$ and $\Phi_{\textrm{ST}}(t_{\textrm{obs}})$ assume the corresponding forms of the critical frequencies, as those are expressed in eq. (\ref{nu_cr}) and Tables \ref{nu_cr-BM-am}, \ref{nu_cr-BM-ma}, \ref{nu_cr-ST-am} and \ref{nu_cr-ST-ma}. The value of $s_{\textrm{t}}$ is given by eq. (\ref{s-crit}) and the relevant entries of Table \ref{s-critical}.

Equation (\ref{break_t}) should also be used to calculate $F_{\textrm{m}}$ at the same observer time as the critical frequencies. The asymptotic expressions are presented in eq. (\ref{FmBM}) and (\ref{FmST}) and are to be evaluated using eq. (\ref{C_pol}) and Table \ref{F_m-table}. Having found the values of all 5 critical parameters at the same observer time, we can now start constructing the spectrum.

\subsection{Shape and flux-normalisation of the spectrum}\label{Spectrum_shape}

In the case of spectrum 1 the characteristic synchrotron frequency $\nu_{\textrm{m}1}$ lies on the optically thin part of the spectrum. Consequently, its value is affected by radiation from the whole blast-wave. In the case of spectrum 2 self-absorption allows only for the front to contribute to the flux close to $\nu_{\textrm{m}2}$. In practice the values of $\nu_{\textrm{m}1}$ and $\nu_{\textrm{m}2}$ are always close to each other (within the same order of magnitude). Therefore, we conclude that the location of $\nu_{\textrm{m}}$ on the spectrum is mostly determined by the conditions at the front and is only slightly affected by what the optical depth of the blast-wave is at that frequency.

On the other hand, values of $\nu_{\textrm{a}1}$ and $\nu_{\textrm{a}2}$ can differ substantially with respect to each other. However, in most cases they will both be either smaller or bigger than $\nu_{\textrm{m}1}$ \textit{and} $\nu_{\textrm{m}2}$. We will be referring to these cases as definite ordering, whereas all other cases will be referred to as indefinite ordering. When $\nu_{\textrm{a}1},\, \nu_{\textrm{a}2} < \nu_{\textrm{m}1},\, \nu_{\textrm{m}2}$  the spectrum will have the form of Fig. \ref{fig_spec_am} (spectrum 1), while if $\nu_{\textrm{m}1},\, \nu_{\textrm{m}2} < \nu_{\textrm{a}1},\, \nu_{\textrm{a}2}$ that of Fig. \ref{fig_spec_ma} (spectrum 2). In the case of indefinite ordering, the actual positions of the two critical frequencies on the spectrum are very close to each other signaling a spectral transition, typically from spectrum 1 to spectrum 2. In terms of observer time, the time-span of this transition is relatively small. Let $t_{\textrm{tr}}$ be the observer time when indefinite ordering sets in. The duration of this transtion (time-span of indefinite ordering) is typically a fraction of $t_{\textrm{tr}}$. During that time the choice of critical frequencies affects the flux across the spectrum by factors of order unity.

In practice, it is preferable to always take the values suggested by both spectra into account, through a consistent weighing method. This way glitches that may appear in the produced lightcurves when switching from one spectrum to another are avoided. Instead, the lightcurves' behaviour smoothly progresses from the early-time spectrum 1 configuration to the late-time spectrum 2. The weight of each spectrum is represented by a power-law dependence in time that contains a characteristic timescale $t_{\textrm{flip}}$ related to the observer time at which the spectrum is transitioning from spectrum 1 to spectrum 2. This characteristic timescale can be estimated numerically by solving eq. (\ref{break_t_ini}) for both $t_{\textrm{T}1}$ (the time at which $\nu_{\textrm{m}1}=\nu_{\textrm{a}1}$) and $t_{\textrm{T}2}$ (the time at which $\nu_{\textrm{m}2}=\nu_{\textrm{a}2}$), and defining

\begin{equation}
t_{\textrm{flip}} = f_{\textrm{flip}} \, \cdot \, \textrm{max}(t_{\textrm{T}1}, t_{\textrm{T}2}).
\end{equation}
We have found that the value of $f_{\textrm{flip}}$ that results in smaller deviations across the parameter space of $[p,k]$ is $1.6$.

The weights of spectrum 1 and 2 can be written as

\begin{equation}
W_1 = \frac{\left(t_{\textrm{obs}}/t_{\textrm{flip}}\right)^{-1}}{\left(t_{\textrm{obs}}/t_{\textrm{flip}}\right)^{-1}\,+\,\left(t_{\textrm{obs}}/t_{\textrm{flip}}\right)},
\end{equation}
\begin{equation}
W_2 = \frac{\left(t_{\textrm{obs}}/t_{\textrm{flip}}\right)}{\left(t_{\textrm{obs}}/t_{\textrm{flip}}\right)^{-1}\,+\,\left(t_{\textrm{obs}}/t_{\textrm{flip}}\right)}.
\end{equation}

The flux at a given frequency will then be

\begin{equation} \label{final_flux}
\textrm{log}\, F = W_1\cdot \textrm{log}\, F_1\, +\, W_2\cdot \textrm{log}\, F_2\, ,
\end{equation}
where $F_1$ and $F_2$ are the fluxes calculated at that frequency through spectrum 1 and 2 (see subsections \ref{spec1_section} and \ref{spec2_section}), respectively.

\subsubsection{Spectrum 1: $\nu_{\textrm{a}} \, < \, \nu_{\textrm{m}}$}\label{spec1_section}

This is the asymptotic case where both $\nu_{\textrm{a}1}$ and $\nu_{\textrm{a}2}$ are smaller than $\nu_{\textrm{m}1}$ and $\nu_{\textrm{m}2}$. Consequently the positions of the critical frequencies are given by $\nu_{\textrm{a}}=\nu_{\textrm{a}1}$ and $\nu_{\textrm{m}}=\nu_{\textrm{m}1}$. The parameters of eq. (\ref{eq-spectrum}) get the following values:
\begin{enumerate}
\item $\nu_0=\nu_{\textrm{a}1}$
\item $\nu_1=\nu_{\textrm{m}1}$
\item $A=F_{\textrm{m}}\left( \nu_0/\nu_1 \right)^{1/3}$
\item $a_1=2$
\item $a_2=1/3$
\item $a_3=(1-p)/2$
\end{enumerate}

\subsubsection{Spectrum 2: $\nu_{\textrm{m}} \, < \, \nu_{\textrm{a}}$}\label{spec2_section}

For spectrum 2 in the asymptotic limit both $\nu_{\textrm{a}1}$ and $\nu_{\textrm{a}2}$ are bigger than $\nu_{\textrm{m}1}$ and $\nu_{\textrm{m}2}$. Thus, the positions of the critical frequencies are given by $\nu_{\textrm{m}}=\nu_{\textrm{m}2}$ and $\nu_{\textrm{a}}=\nu_{\textrm{a}2}$. However, in spectrum 2 the flux at $\nu_{\textrm{m}}$ is not $F_{\textrm{m}}$; that would be the case if it were not for absorption. We can use this fact to first calculate the flux at $\nu_{\textrm{a}}$. Although in the present spectrum-configuration the actual position of $\nu_{\textrm{m}}$ is given by $\nu_{\textrm{m}2}$, it is $\nu_{\textrm{m}1}$ that we should use for obtaining the flux at $\nu_{\textrm{a}}$. The variables become:
\begin{enumerate}
\item $\nu_0=\nu_{\textrm{m}2}$
\item$\nu_1=\nu_{\textrm{a}2}$
\item $A=F_{\textrm{m}} \left( \frac{\nu_{\textrm{a}2}}{\nu_{\textrm{m}1}} \right)^{(1-p)/2} \left( \frac{\nu_{\textrm{m}2}}{\nu_{\textrm{a}2}} \right)^{2.5}$
\item $a_1=2$\item $a_2=2.5$
\item $a_3=(1-p)/2$
\end{enumerate}

\subsection{Sharpness parameters of spectral breaks}\label{sharp_spectral}

The only two parameters left to specify in eq. (\ref{eq-spectrum}) are $s$ and $h$, the sharpness of the first and the second break in the spectrum, respectively. In order to assign the proper sharpness to each break we first have to compare $t_{\textrm{obs}}$ to $t_{\textrm{NR}}$ (see eq. \ref{tnr}). There are four distinct cases, as illustrated in Fig. \ref{evo-s}:

\subsubsection{$t_{\textrm{obs}}<t_{\textrm{i}}$}

In this case all sharpness parameters attain their BM values as these are given in Table \ref{s-rel}.

\subsubsection{$t_{\textrm{obs}}>t_{\textrm{f}}$}

In this case all sharpness parameters attain their ST values as these are given in Table \ref{s-nonrel}.

\subsubsection{$t_{\textrm{i}}<t_{\textrm{obs}}<t_{\textrm{NR}}$}

In this case breaks $\nu_{\textrm{m}2}$ and $\nu_{\textrm{a}2}$ retain their BM sharpness. The other two exhibit some evolution towards the corresponding ST values. Namely, the sharpness around $\nu_{\textrm{m}1}$ will be

\begin{equation}
s_{\textrm{m}1} = 0.09 \, \, \textrm{log}\! \left( \frac{t_{\textrm{i}}}{t_{\textrm{obs}}} \right) + s_{\textrm{i}},
\end{equation}
where $s_{\textrm{i}}$ is the sharpness of the particular break in the BM regime.

The sharpness around $\nu_{\textrm{a}1}$ will be

\begin{equation}\label{sa1-evo}
s_{\textrm{a}1} = \frac{(s_{\textrm{f}}-s_{\textrm{i}})}{3} \,\, \textrm{log} \! \left( \frac{t_{\textrm{obs}}}{t_{\textrm{i}}} \right) + s_{\textrm{i}},
\end{equation}
where $s_{\textrm{i}}$ and $s_{\textrm{f}}$ are the sharpness parameters at the BM and ST phases, respectively.

\subsubsection{$t_{\textrm{NR}}<t_{\textrm{obs}}<t_{\textrm{f}}$}

In this final case all breaks exhibit a sharpness evolving towards its ST value. For $\nu_{\textrm{m}1}$ the sharpness will be given by

\begin{equation}
s_{\textrm{m}1} = (s_{\textrm{f}}-s_{\textrm{i}}+0.18) \,\, \textrm{log} \! \left( \frac{t_{\textrm{obs}}}{t_{\textrm{f}}} \right) + s_{\textrm{f}},
\end{equation}
while for $\nu_{\textrm{a}1}$ the value of the sharpness is still given by eq. (\ref{sa1-evo}).

The sharpness around $\nu_{\textrm{a}2}$ will be given by

\begin{equation}
s_{\textrm{a}2} = (s_{\textrm{f}}-s_{\textrm{i}}) \,\, \textrm{log} \! \left( \frac{t_{\textrm{obs}}}{t_{\textrm{NR}}} \right) + s_{\textrm{i}},
\end{equation}
while for $\nu_{\textrm{m}2}$ we find a similar result

\begin{equation}
s_{\textrm{m}2} = (s_{\textrm{f}}-s_{\textrm{i}}) \,\, \textrm{log} \! \left( \frac{t_{\textrm{obs}}}{t_{\textrm{NR}}} \right) + s_{\textrm{i}}.
\end{equation}

\subsection{Examples of results}

We have described a practical implementation of our results to construct spectra at any given time, based on values for the physical quantities characterising the burst. We now show comparisons between simulation-generated spectra and spectra that have been constructed using the provided flux prescriptions.

\begin{figure}
\includegraphics[width=\columnwidth]{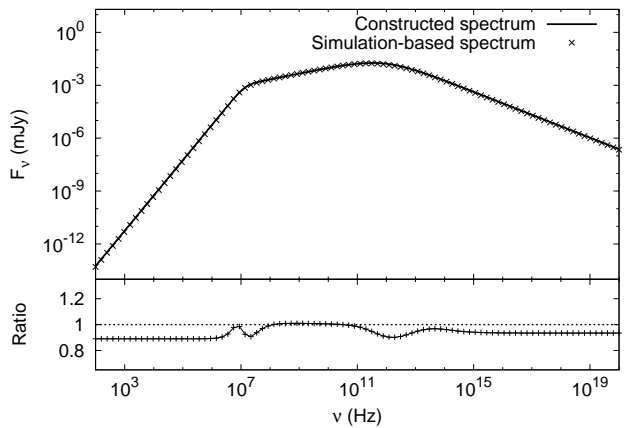}
\caption{A typically good match between an analytically constructed spectrum and one based on a simulation. Both are taken at $100$ days. $\nu_{\textrm{a}}$ lies at $\sim 10^7\, \textrm{Hz}$ and $\nu_{\textrm{m}}$ at $\sim 10^{12}\, \textrm{Hz}$. Model parameters for both spectra are: $E_{52}=1,\, n_0=1,\, p=2.3,\, k=0,\, \xi=10^{-2},\, \epsilon_{\textrm{e}}=10^{-1},\, \epsilon_{\textrm{B}}=10^{-2},\, d_{28}=1,\, z=0.56$.}
\label{fig-fit-1}
\end{figure}

In Fig. \ref{fig-fit-1} a comparison between a simulation-based spectrum and an analytic one is shown. In all power-law segments and the linking breaks, the flux-prediction is never more than $~10\%$ off compared to the simulation-based data. We can translate these deviations into relative errors for the values of the physical parameters. This we do by adjusting their values so that those deviations vanish in particular regimes of the spectrum. For the blast-wave energy ($E_{52}$) the error ranges from $3\%$ up to $15\%$ depending on which power-law segment (or spectral break) one uses for the comparison. In the case of $n_0$ the maximum error is $15\%$. For $p$ the difference is of order $0.05$, while for $k$ it is of order $0.08$. For $\epsilon_{\textrm{e}}$ and $\epsilon_{\textrm{B}}$ the error is of the order $10\%$ while for $\xi$ it reaches up to $30\%$.

\begin{figure}
\includegraphics[width=\columnwidth]{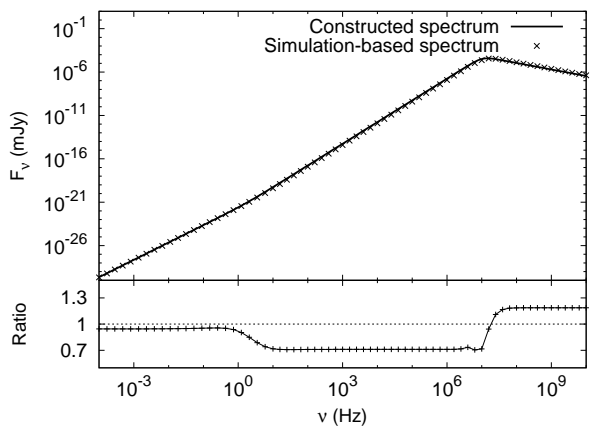}
\caption{An example of a constructed spectrum that shows relatively large deviations from the numerical result in the optically thin part of the spectrum. Both spectra are taken at $500$ days. $\nu_{\textrm{m}}$ lies at $\sim 1\, \textrm{Hz}$ and $\nu_{\textrm{a}}$ at $\sim 10^{7}\, \textrm{Hz}$. Model parameters are: $E_{52}=1,\, n_0=1,\, p=2.5,\, k=0.5,\, \xi=1,\, \epsilon_{\textrm{e}}=10^{-4},\, \epsilon_{\textrm{B}}=10^{-2},\, d_{28}=1,\, z=0.56$.}
\label{fig-fit-2}
\end{figure}

In Fig. \ref{fig-fit-2} we present another comparison between a simulation-based spectrum and a constructed one. This one was chosen for exhibiting one of the largest deviations we have encountered. While the self-absorbed part of the spectrum is matched well by the constructed spectrum, flux in the $\nu^{1-p/2}$ segment differs by $\sim 25\%$. The corresponding errors in the derivation of values for physical quantities are the following: for $E_{52}$ up to $16\%$, for $n_0$ $10\%-90\%$ (flux in the $\nu^{1-p/2}$ segment depends very weekly on $n_0$ for these model parameters), for $\xi$ up to $35\%$, for $\epsilon_{\textrm{e}}$ $15\%$, for $\epsilon_{\textrm{B}}$ $30\%$, while for $p$ and $k$ we find differences up to $0.1$ and $0.2$, respectively.

We stress that these deviations are not with respect to a best-fit value but are indicative of how much every parameter should be tweaked to match fluxes in individual power-law segments of the spectrum. More often than not such a tweak would actually produce a rather bad fit overall. Thus the deviations we have listed may be viewed as an upper limit to what a broadband fit would produce.

\subsection{Application to mildly relativistic outflows}\label{application}

Recently \citet{Nakar2011} have discussed the radio-signal following the ejection of spherical, Newtonian or mildly relativistic outflows expected from binary neutron star mergers. They estimate that due to the low initial Lorentz factors of these outflows, their deceleration (and entry to the ST phase) will be manifested at $t_{\textrm{dec}}\sim 60\, \textrm{days}$ observer time, for $E_{52}=0.01$, $n_0=1$, $k=0$, $\beta_{\textrm{i}}\sim 1$ (initial velocity). This is also the time at which optically thin emission at $\nu_{\textrm{obs}}=5\,\textrm{GHz}$ will peak in the range $0.01-0.1\, \textrm{mJy}$, for a distance of the source in the range $1-3\,\textrm{Gpc}$. Here, we test these estimates using the prescriptions presented in this paper.

The model we have developed in this study is based on (and therefore, applicable to) outflows that are initially ultrarelativistic. Thus, it is not obvious that it can be used to model non-relativistic outflows. Order of magnitude calculations in the lab frame can illustrate the limitations. A relativistic outflow of coasting Lorentz factor $\Gamma_{\textrm{i}}$ will slow down after sweeping mass $\Gamma_{\textrm{i}}$ times smaller than the mass of the ejecta (\citealt{Rees1992}). This will happen at a time

\begin{equation}
t_{\textrm{BM}} = \left(  \frac{3E}{4 \pi \rho_1 c^5}  \right)  ^{1/3} \, \Gamma_{\textrm{i}}^{-2/3}.
\end{equation}
From that point onwards the outflow will decelerate according to the BM solution ($\Gamma \propto t^{-3/2}$) becoming Newtonian ($\Gamma \sim 1$) at
\begin{equation}\label{t_ST_rel}
t_{\textrm{N}} = \left(  \frac{3E}{4 \pi \rho_1 c^5}  \right)  ^{1/3}.
\end{equation}
The corresponding radius is
\begin{equation}\label{r_ST_rel}
r_{\textrm{N}} = t_{\textrm{N}} \, c.
\end{equation}
Equations (\ref{t_ST_rel}) and (\ref{r_ST_rel}) effectively mark the onset of the ST phase.

In the case of sub- and mildly relativistic outflows the deceleration time (also marking the transition to the ST phase) occurs when the swept mass is comparable to the rest mass of the ejecta

\begin{equation}\label{t_ST_nonrel}
t_{\textrm{dec}} = \left(  \frac{3E}{2 \pi \rho_1 c^5}  \right)  ^{1/3} \, \beta_{\textrm{i}}^{-5/3}.
\end{equation}
At $t_{\textrm{dec}}$ the shock is at a radius 
\begin{equation}\label{r_ST_nonrel}
r_{\textrm{dec}} = \beta_{\textrm{i}} \, t_{\textrm{dec}} \, c.
\end{equation}

From eq. (\ref{t_ST_rel})-(\ref{r_ST_nonrel}) it is clear that as $\beta_{\textrm{i}} \rightarrow 1$ the onset of the ST phase for Newtonian outflows approaches that of the relativistic analog with the same energy. This implies that fast ($v\sim c$) outflows (regardless of the Lorentz factor) have no \textit{memory} of their history from $t_{\textrm{dec}}$ onwards. Therefore we can apply the ST scalings of the flux-presciptions to a mildly relativistic outflow (as the one considered by \citealt{Nakar2011}) at observer times $t_{\textrm{obs}} \geq t_{\textrm{dec}}$. In the sub-relativistic case ($\beta_{\textrm{i}} \ll 1$) eq. (\ref{t_ST_nonrel}) and (\ref{r_ST_nonrel}) imply that the outflow will decelerate later and at a greater radius. Nevertheless, the ST scalings of the flux-prescriptions apply at observer times $t_{\textrm{obs}} \gg t_{\textrm{dec}}$, i.e. sufficiently later than the deceleration time.

 For the application to mildly relativistic outflows we have set $E_{52}=0.01,\, n_0=1,\,k=0$ for the macroscopic parameters of the blast-wave and its environment and $\xi=1,\,\epsilon_{\textrm{e}}=\epsilon_{\textrm{B}}=0.1$, for the microphysics, while $\nu_{\textrm{obs}}=5\,\textrm{GHz}$ and $t_{\textrm{obs}}=60\,\textrm{days}$. The electron spectral index $p$ is varied within the range $2.1-3.0$, while for the distance we have taken the two extreme values of $1$ and $3\,\textrm{Gpc}$.

\begin{figure}
\includegraphics[width=\columnwidth]{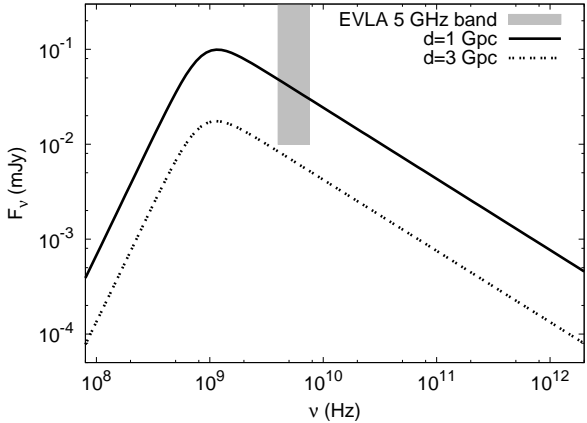}
\caption{Synchrotron spectra of a mildly relativistic outflow at distances of $1$ and $3\, \textrm{Gpc}$, taken at 60 days observer time. The grey shaded area represents the $5\, \textrm{GHz}$ band of the EVLA with a bandwidth of $3.5\, \textrm{GHz}$ and a $4\sigma$ detection threshold of $\sim 10\, \mu\textrm{Jy}$ for $1\, \textrm{hr}$ integration (\citealt{Perley2011}). The value of $p$ for both spectra is 2.5. The other physical parameters have the following values: $E_{52}=0.01,\, n_0=1,\,k=0,\, \xi=1,\,\epsilon_{\textrm{e}}=\epsilon_{\textrm{B}}=0.1$.}
\label{fig_NS-NS}
\end{figure}

In accordance with \citet{Nakar2011} we find that $\nu_{\textrm{obs}}$ is in the optically thin part of the spectrum for all cases. In this regime the flux increases monotonically for an increasing $p$ and for $p=3.0$ it is about four times higher than the $p=2.1$ case. For $d_{28}=0.31\,(\sim1\,\textrm{Gpc})$ the flux at $5\,\textrm{GHz}$ lies in the range $0.015-0.06\,\textrm{mJy}$, depending on the value of $p$. At $d_{28}=0.93\,(\sim3\,\textrm{Gpc})$ we find the flux to be always below $0.01\, \textrm{mJy}$, albeit marginally for relatively high values of $p$. This is illustrated in Fig. \ref{fig_NS-NS} where two spectra are shown corresponding to distances of $1$ and $3\, \textrm{Gpc}$. They are both taken at $t_{\textrm{obs}}=60\,\textrm{days}$, for a characteristic value of $p=2.5$. 

Binary neutron star mergers are believed to be the progenitors of short GRBs (see \citealt{Nakar2007} and references therein) and are observed in a variety of environments, like elliptical, spiral and irregular galaxies (\citealt{Berger2009}). A considerable fraction of them, however, appear to be host-less, occuring in the intergalactic medium and thus surrounded by a much more tenuous gas than that commonly found inside galaxies (\citealt{Berger2010}). We have repeated the calculation of the spectrum from a spherical outflow resulting from a NS-NS merger for a surrounding medium of density $n_0=10^{-3}\, \textrm{cm}^{-3}$, where the lower density results in a later onset of the ST phase at $\sim 400\, \textrm{days}$. Keeping all other parameters constant we find that the radio signal at $5\, \textrm{GHz}$ will be detectable by the EVLA up to a distance of $\sim 100\, \textrm{Mpc}$.

The implication of these results is that moderately energetic outflows ($E=10^{50}\, \textrm{erg}$) expected to accompany NS-NS mergers (\citealt{Rezzolla2011}) can produce synchrotron radiation detectable by the EVLA from distances up to $\sim 1 \, \textrm{Gpc}$, larger than the detection horizon of the upcoming versions of gravitational-wave detectors (\citealt{Nakar2011}). This assumes that the density of the matter surrounding the merger is of the order $1\, \textrm{cm}^{-3}$. These radio signals will peak at timescales of the order of a few months, if the corresponding outflows have initial velocities close to the speed of light. In the case of more tenuous circumburst media the ST timescale grows and the detection horizon of the EM signal drops accordingly. The presented flux-prescriptions are applicable throughout the ST phase of these outflows.


\section{Discussion}\label{discuss}
We present analytic flux-prescriptions, for broadband synchrotron spectra originating from GRB outflows, suitable for fast and detailed modelling of the afterglow phase. They are applicable throughout the evolution of observed afterglows, during which external shocks are the dominant source of particle acceleration, and account for the exact shape of the synchrotron spectrum, including self-absorption, but ignoring cooling. These prescriptions are based on high-resolution, one-dimensional, hydrodynamic simulations performed using the adaptive mesh refinement code \textsc{amrvac}. To obtain spectra we have employed a radiation code that solves the equation of radiative transfer through the evolving blast-wave as this is determined by the simulations. The presented formulas carry two components. The first is derived analytically and expresses the dependence of the flux on relevant physical parameters ($E_{52},\, n_0,\, p,\, k,\, \xi,\, \epsilon_{\textrm{e}},\, \epsilon_{\textrm{B}}, \nu_{\textrm{obs}},\, t_{\textrm{obs}},\, d_{28},\, z$), while the second component reflects the calibration that the results of simulations have introduced to the flux-levels.

For each asymptotic dynamical regime (BM and ST) we provide prescriptions for the flux at every power-law segment but also at frequencies close to spectral breaks. These are modelled as smoothly broken power-laws with the sharpness of the break given in terms of the structure of the surrounding medium ($k$) and the electron distribution ($p$). During the transrelativistic regime we find that the values of critical frequencies ($\nu_{\textrm{m}}$, $\nu_{\textrm{a}}$) and peak flux ($F_{\textrm{m}}$) of the synchrotron spectrum show a gradual transition from the asymptotic power-law behaviour in the BM phase to the corresponding one in the ST phase. This fact has allowed us to model their temporal profiles as smoothly broken power-laws. For every parameter ($F_{\textrm{m}}$, $\nu_{\textrm{m}}$ and $\nu_{\textrm{a}}$) we provide formulas describing the sharpness of these breaks in terms of $p$ and $k$. In order to model the evolution of a spectral break's sharpness, we have recognized the unique pattern that each break exhibits. We have introduced $t_{\textrm{NR}}$ whose derivation is based on considerations of the outflow dynamics. The result is a set of analytic expressions that extend the applicability of the flux-prescriptions to any given observer time.

An element of this study worth emphasising is the inclusion of $k$ (representing the structure of the circumburst medium) as a fitting parameter. This is motivated by the fact that environments of stars with variable mass-loss rates (such as massive stars, prime candidates for long GRB progenitors) can have structures more complex than the usually assumed $k=0\textrm{ or }2$ (\citealt{Ramirez-Ruiz2005}) and fits using $k$ as a free parameter do not exclude such a possibility \citep{Yost2003, Curran2009}. As can be seen in Tables \ref{F-rel-g} and \ref{F-nonrel-g} the impact of $k$ on flux values is modest and varies smoothly across a plausible range of $k$-values $[0,2]$. Nevertheless, its effect is measurable in light of the provided formulas, contributing an extra tool to afterglow fitting and addressing the nature of GRB progenitors.

Beyond the context of GRBs, the provided prescriptions are useful for modelling synchrotron emission from spherical adiabatic blast-waves of arbitrary velocity (with the limitations analysed in section \ref{application}) as long as they have swept up enough mass to be decelerating. Obvious applications include type Ibc supernovae (\citealt{Soderberg2010}), often associated with GRBs (\citealt{Woosley2006}) and mildly- or sub-relativistic spherical outflows from binary neutron star mergers. The latter are candidates for providing the EM counterpart (peaking at radio frequencies) to a possible signal of gravitational waves (\citealt{Metzger2012b}). By applying the ST scalings of the presented flux-prescriptions on mildly relativistic outflows we show that prospects of detecting such radio signals from within the horizon of gravitational wave detectors, LIGO (\citealt{Abbott2009}) and Virgo (\citealt{Acernese2008}), are realistic (\citealt{Nakar2011}).

\vspace{0.4cm}

It is interesting to note that apart from the dependence on $p$ and $k$, we have also found that the sharpness of a spectral break can be influenced to some extent by the microphysical parameters $\epsilon_{\textrm{e}}$, $\epsilon_{\textrm{B}}$ and $\xi$. This has been particularly seen in spectral breaks that involve absorption. The reason for this is the dependence of the absorption coefficient on the chosen microphysics through eq. (\ref{abs-eq}). The microphysical parameters in effect regulate the physical depth of the blast-wave corresponding to a given value of the optical depth. Therefore an increase/decrease of $\alpha'_{\nu}$ results in a less/more diverse sample of local electron  distributions contributing to the flux across a spectral break and thus a sharper/smoother transition. We have chosen not to include the effect of the microphysics on the sharpness-formulas, as it typically influences $s$ by no more than $10-15\%$ (the flux to a lesser extent) and it would greatly complicate the heuristic equations we present.

Contrary to the approach on the evolution of a spectral break's sharpness, where the introduction of $t_{\textrm{NR}}$ is useful, when describing the temporal evolution of the spectrum's critical parameters we deliberately choose not to use such a timescale. The reason is that, as it turns out, there is no such thing as a single global timescale applicable to the behaviour of all observable quantities. Instead, every critical parameter of the spectrum is chracterised by its own break time, the meeting point of the BM and ST asymptotes. One can verify that by computing $t_0$ of eq. (\ref{break_t_ini}) for a few critical parameters of the same model, by equating the asymptotic expressions. They will be found to differ by factors up to a few. This happens because at any given observer time all these parameters are affected by contributions of radiation from various parts of the outflow, emitted within a range of lab-frame times. For each parameter the weight of these contributions will differ, leading to the inference of contrasting timescales by an observer. This stresses the need for models that can naturally account for this kind of features, by implementing accurate calculations of the blast-wave dynamics and the shape of the spectrum.

\begin{figure}
\includegraphics[width=\columnwidth]{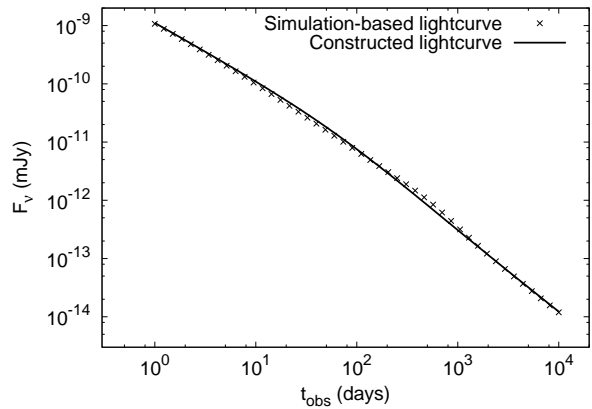}
\caption{Simulation-based lightcurve taken at $10^{20}\, \textrm{Hz}$ (no cooling taken into account). A bump at $\sim 500$ days is apparent. For comparison we have plotted the constructed lightcurve based on the presented flux prescriptions. Model parameters are: $E_{52}=1,\, n_0=1,\, p=2.3,\, k=0,\, \xi=1,\, \epsilon_{\textrm{e}}=10^{-4},\, \epsilon_{\textrm{B}}=10^{-2},\, d_{28}=1,\, z=0.56$.}
\label{thin_light}
\end{figure}

By inspection of Fig. \ref{fig_F_m_fit} one can realise that the broken power-law approach is an approximation to the actual behaviour of any critical spectrum parameter during the transrelativistic phase. The parameter that exhibits the strongest deviation from this description is $\nu_{\textrm{a}2}$. The reason for this can be traced to the behaviour of the flux in the optically thin part of the spectrum. An example of this behaviour is shown in Fig. \ref{thin_light}, where a feature readily apparent is a smooth bump centered at $\sim 500$ days (this can also be seen in Fig. 10 of \citealt{vanEerten2010a}). This introduces a similar feature in the temporal profile of $\nu_{\textrm{a}2}$. As a result, the actual values of that frequency can deviate as much as $\sim 15\%$ from the fitting function (smooth power-law break) at observer times relatively close to $t_{\textrm{NR}}$. This can have an effect on constructed lightcurves if the self-absorbed part of spectrum 2 is used and only during the transrelativistic phase. The impact on flux levels is stronger than the deviations shown in Fig. \ref{thin_light} because the flux at the $\nu^{2}$ and $\nu^{5/2}$ segments of spectrum 2 scales as $\nu_{\textrm{a}2}^{-(p+4)/2}$. We therefore recommend using eq. (\ref{closure}) in all cases as this method provides a more accurate and consistent way of constructing spectra and lightcurves.

\vspace{0.4cm}

By now there are a number of studies in the literature that present formulas calculating spectra from GRB afterglows. The importance of taking into account the blast-wave structure and the exact shape of the spectrum has been stressed by discrepancies in the derived values of physical parameters between simple \citep{Wijers1999} and more elaborate \citep{Granot2002,vanEerten2009} models. Inclusion of details regarding the shock structure can be done either analytically (in either of the two asymptotic regimes of the dynamics) or through the performance of simulations, as is done in the present work. The advantage of numerical simulations is that they cover the transrelativistic phase of the outflow and the level of detail they provide in all cases. The disadvantage is the price they come at, both in terms of time and resources.

In this paper we provide an efficient way of utilizing the benefits of simulations, as those are reflected on the presented analytic prescriptions. A similar approach has been taken by \citet{vanEerten2012a}, who are basing their fitting method on the scale-invariance of light-curves. This method naturally accounts for features like sideways spreading of the jet and off-axis observation angles, features that only arise in simulations of at least two dimensions and cannot be captured in the context of this research. However, it requires the use of a large database of light curves which does not yet exist. So far no study of the afterglow radiation using 2D hydrodynamic simulations \citep{Zhang2009, Wygoda2011, DeColle2012b} has resulted in the derivation of flux prescriptions. In fact this is the first time, even for the simple spherical case, that simulation-based flux prescriptions beyond the BM phase are presented. The box-fit method of \citet{vanEerten2012b} does provide a fitting code but requires the use of a parallel computer network in order to fit data by iterating through a ``box'' of simulations. Therefore, analytic flux-prescriptions based on 1D simulations, as the ones presented here, can be the base for comparisons with future work in that direction based on 2D simulations. Moreover, 1D models are always relevant both at observer times before the jet break (when most parts of the outflow are causally disconnected) and at late times when the outflow is roughly spherical and allow for accurate calorimetry of jetted outflows after the jet-break but well before spherical symmetry has been reached, as long as the observer is not far off-axis (\citealt{Wygoda2011}).

\vspace{0.4cm}

In all the simulations that we have performed to arrive at the presented flux prescriptions, the microphysical parameters have been kept constant throughout the run and at every part of the outflow. This is by no means guaranteed and therefore introduces an uncertainty in our results. An interesting topic for further study is the implementation of evolving microphysical parameters and their effect on the flux prescriptions. Such an evolution is expected on theoretical grounds for some of those parameters (\citealt{Granot2006}). A qualitative study of the evolution of $\xi$ at the shock front and the evolution of $\epsilon_{\textrm{B}}$ downstream has already been presented in \citet{vanEerten2010a}. Meanwhile, there is also growing amount of observational evidence for this process taking place in GRB afterglows \citep{Panaitescu2006, Kong2010, Filgas2011}. Incorporating effects like time-dependence of the microphysics into flux prescriptions can extend the predictions of the standard fireball model and thus broaden the theoretical framework within which observations are currently being interpreted.

\section{Conclusions}

We have used high-resolution 1D hydrodynamic simulations to calibrate flux scalings of synchrotron, self-absorbed radiation for GRB afterglows in the relativistic and Newtonian dynamical phases (BM and ST, respectively). The transition from the former to the latter is well described by approximating the evolution of spectral parameters (maximum flux and positions of critical frequencies) by power-law breaks connecting the two asymptotic behaviours. The properties of these breaks have been modelled in terms of the values of the physical parameters describing the blast-wave. This way we have managed to encapsulate the precision of the performed simulations into a set of analytic formulas that trace the full evolution of GRB afterglows, from the ultrarelativistic to the Newtonian phase. Due to the general nature of the prescriptions, they are applicable to any source characterised by emission of synchrotron radiation from an adiabatic blast-wave.

A numerical code containing a practical implementation of the results presented in this paper combined with a fitting code is freely available on request and on-line at http://www.astro.uva.nl/research/cosmics/gamma-ray-bursts/software/.

\section{Acknowledgements}

This research was supported by NOVA and in part by NASA through grant NNX10AF62G issued through the Astrophysics Theory Program and by the NSF through grant AST-1009863. RAMJ acknowledges support from the ERC via Advanced Investigator Grant no. 247295. We thank SARA Computing and Networking Services (www.sara.nl) for their support in using the Lisa Compute Cluster. We thank Evert Rol and Alexander van der Horst for useful discussions.

\label{lastpage}

\end{document}